\documentclass{article}
\usepackage{arxiv}

\usepackage[utf8]{inputenc} 
\usepackage[T1]{fontenc}    
\usepackage{hyperref}       
\usepackage{url}            
\usepackage{booktabs}       
\usepackage{amsfonts}       
\usepackage{nicefrac}       
\usepackage{microtype}      
\usepackage{lipsum}
\usepackage{multirow}
\usepackage{amssymb,amsmath,epsfig}
\newcommand{\beq}{\begin{equation}}
\newcommand{\eeq}{\end{equation}}
\newcommand{\bea}{\begin{eqnarray}}
\newcommand{\eea}{\end{eqnarray}}

\begin{document}
\title{Particle dynamics around a dyonic charged black hole}

\author{{Saeed Ullah Khan\thanks{saeedkhan.u@gmail.com}} \, and \, {Jingli Ren\thanks{Corresponding author: renjl@zzu.edu.cn}}\vspace{0.2cm} \\\vspace{0.08cm}
Henan Academy of Big Data/School of Mathematics and Statistics, Zhengzhou University,\\
Zhengzhou 450001, China.}
\date{}
\maketitle
\begin{abstract}
In this article, we study the circular motion of particles and the well-known Penrose mechanism around a Kerr-Newman-Kasuya black hole spacetime. The inner and outer horizons, as well as ergosurfaces of the said black hole, are briefly examined under the effect of spin and dyonic charge. Moreover, by limiting our exploration to the equatorial plane, we discuss the characteristics of circular geodesics and investigate both photons, as well as marginally stable circular orbits. It is noted that black hole charge diminishing the radii of photon and marginally stable circular orbits. To investigate the nature of particle dynamics, we studied the effective potential and Lyapunov exponent. While inspecting the process of energy extraction, we derived the Wald inequality, which can help us to locate the energy limits of the Penrose process. Furthermore, we have found expressions for the negative energy states and the efficiency of energy extraction. The obtained result illustrates that both black hole rotation and dyonic charge contributes to the efficiency of energy extraction.
\end{abstract}

\section{Introduction}
\label{sec:1}
The physics of a black hole (BH), is perhaps among the most perplexing and still a questionable factor of Einstein's theory of relativity. Due to its extremely dense spacetime region, it consists of abundantly powerful gravitational attraction. Henceforth, pulling its surrounding matter is the intrinsic property of a BH, known is the accretion \cite{Cunha}. BHs are probably the main source of jets and gamma-ray bursts; quasars; accretion disc and gravitational bending of light \cite{Berti}. As a result, the process of energy extraction from a BH is of ample interest in astrophysics. Since the exact form of the process is still lacking, hence to understand the central mechanism, the phenomena occurs within ergoregion of a BH is very important \cite{Frolov}. Moreover, the ergoregion of a BH is also momentous, due to the Hawking radiation \cite{Hawking} as they can be observed in this region.

Although there is no clear assurance to the existence of gravitomagnetic charges, it is still accepted that there may exist an originating tool of such charges in gravitational interactions. The gravitomagnetic charge is the proposed gravitational analogue of Dirac's magnetic monopole \cite{Chakraborty}. Nevertheless, if it exists in the universe, it could cause a remarkable outcome in the field of astrophysics. Notwithstanding,  it is notable that dyon is a pole having both of the electric, as well as magnetic charges. The Kerr-Newman-Kasuya (KNK) spacetime \cite{Kasuya} expresses a 4-dimensional BH, possess both of the electric and magnetic charges. Earth's orbit is the optimum spot for the determination of gravitomagnetism. Similar to a rotational sphere of electric charge that generates a magnetic field, Earth is also assumed to generate a gravitomagnetic field. Moreover, the outcomes of modern gauge theories also support the existence of monopole \cite{Mavromatos,Chanyal}.

Particle dynamics around BH spacetimes have been widely studied by many researchers \cite{Bardeen,Detweiler,Mashhoon,Shibata,Pretorius,Stuch5,Kolos1,TOteeva,Khan1}. Since, geodesics are of great interest, as they could convey eminent information and reveal the prolific structure of background geometry. It's examination also assist us to learn about the orbiting gravitation field of spacetime. Among different types of geodesic motion, the circular one is much engaging. Chandrasekhar \cite{Chandrasekhar} was perhaps the first to explore the geodesics of Schwarzschild, Reissner-Nordstr{\"o}m (RN) and Kerr BHs. Besides, circular geodesics are also useful for understanding and explaining the quasinormal modes of a BH \cite{Nollert}. Pugliese et al. \cite{Pug1,Pug2} by investigating the spatial regions of particle dynamics around a RN spacetime figured out the situation of circular orbits, with the vanishing angular momentum. Hod \cite{Hod} by studying the hairy BH, extrapolates that non-trivial response of the hair should expand over the circular null geodesic. The circular geodesic motion of photon and test particle of Bardeen and Ayon-Beatu-Garcia geometry illustrates a spherically symmetric non-singular BH or no-horizon spacetimes \cite{Stuch3}. By exploring energy bound in ergoregion, it is shown that no extreme BH can exist with a critical value of the inherent angular momentum of BH rotation \cite{Grib}. The geodesics of KN anti-dS BH is investigated and found that the stable circular orbit will exist if $r \in \left(0, r_{H-}\right)$ \cite{Shanjit}. Sharif and Shahzadi \cite{Sharif1} by studying the circular geodesics, extrapolates that Kerr-MOG (modified theory of gravity) BH has more stable circular orbits, as compared to Kerr and Schwarzschild BHs.

In the year of 1969, Penrose \cite{Penrose1} and then Penrose and Floyd  \cite{Penrose2} initiated an interesting and precise energy extraction mechanism for rotating BHs. Maximum efficiency of the said mechanism is found to be 20.7$\%$, for the extreme Kerr BH. Bhat et al. \cite{Bhat} figured out that, the magnetic Penrose mechanism could be remarkably efficient and its efficiency may exceed than the maximum efficiency of $100\%$ \cite{Dadhich}. The study of Penrose mechanism was further broadened to the higher dimensional BHs and black rings \cite{Nozawa}; to the five-dimensional spinning supergravity BH \cite{Prabhu}. Lasota et al. \cite{Lasota} found that rotating energy can be extracted from any type of matter satisfying the condition of weak energy, if and only if, it can absorb its angular momentum and negative energy. Mukherjee \cite{Mukherjee} by studying the Penrose mechanism, deduced that more energy can be extracted in the case of Kerr BH in comparison with the non-rotating case. 

In Penrose mechanism, the efficiency of energy extraction can be defined as the ratio of energy gain to the input energy. It is observed that in the case of non-Kerr BH, the efficiency of energy extraction increases with the increase of deformation parameter \cite{Liu}. The evaluation of Penrose mechanism around a non-singular rotating BH, reveals that BH electric charge diminishing the efficiency of energy extraction \cite{Toshmatov}. While in the case of Kerr-MOG BH, it is the rotational parameter which enriches the efficiency of energy extraction \cite{Shahzadi}. Recently, Khan et al. \cite{Khan} studied the braneworld Kerr BH and concluded that both spin and brane parameter of the BH, contributes to the efficiency of Penrose mechanism.

This paper explores circular geodesics and the Penrose mechanism around a KNK spacetime. In the following section, we reviewed the KNK spacetime in detail, together with its horizons and both of the inner and outer ergosurfaces. Notation and notions used in the current work are introduced in section  \ref{sec:2}. In section \ref{sec:3}, we investigate particle dynamics together with the effective potential and Lyapunov exponent. The main objective of section \ref{sec:4}, is to study the Penrose mechanism within ergosphere of a KNK BH. The negative energy states and efficiency of the energy extraction are also studied in section \ref{sec:4}. 
Finally, in the last section (\ref{sec:5}), we provide discussions and a thorough summary of our obtained results.
\section{Kerr-Newman-Kasuya spacetime}
\label{sec:2}
According to the exact solution of a spinning dyonic BH, the KNK BH spacetime in Boyer-Lindquist coordinates takes the form \cite{Kasuya,Ovgun}
\begin{equation}\label{e1}
ds^2=g_{tt}dt^2 + 2 g_{t\phi} {dtd\phi} + g_{\phi \phi} d\phi^2 + g_{rr}dr^2+ g_{\theta \theta}{d\theta^2}.
\end{equation}
With
\begin{eqnarray}\nonumber
&&g_{tt}=-\left(\frac{\Delta-a^2\sin^2\theta}{\Sigma} \right), \quad\, g_{rr}=\frac{\Sigma}{\Delta}, \quad\,  g_{\theta \theta}=\Sigma , \\\label{coeff}
&&g_{\phi \phi}= \left(\frac{(a^2+r^2)^2-\Delta a^2\sin^{2}\theta}{\Sigma}\right)\sin^{2}\theta,\\\nonumber
&&g_{t\phi}= -\left(\frac{a^2+r^2-\Delta}{\Sigma}\right)a{\sin^{2}\theta},
\end{eqnarray}
and
\begin{equation}\label{delta}
\Delta= r^2-2Mr +e{^2}+Q{^2}+a^2, \quad \Sigma=r^2+a^2 \cos^{2}\theta \,.
\end{equation}
In which $M$, $e$ and $Q$, respectively represent the mass, electric and magnetic charges, whereas the parameter $a$ denotes spin of the BH. The metric \eqref{e1}, reduces to the KN BH by substituting $Q=0$; to the Kerr BH if $e=Q=0$ \cite{Kerr}; to the RN BH if $a=Q=0$; and to the Schwarzschild BH by letting $a=e=Q=0$ \cite{Schwarz}. Similarly to that of Kerr BH, the metric \eqref{e1} has exactly two horizons, namely the event and Cauchy horizons. The event and Cauchy horizons and both of the inner and outer ergosurfaces can be obtained, respectively by solving $\Delta=0$ and $g_{tt}=0$
\begin{eqnarray}\label{eh}
&r_{H\pm}=M \pm \sqrt{ M^2-e{^2}-Q{^2}-a^2},\\\label{sls}
&r_{es\pm}=M \pm \sqrt{ M^2-e{^2}-Q{^2}-a^2 \cos^2 \theta}.
\end{eqnarray}
The above Eqs. \eqref{eh} and \eqref{sls} should respectively satisfy the constraints of
\begin{eqnarray}\nonumber
M^2 \geq  e{^2}+Q{^2}+a^2 \quad \text{and} \quad   M^2 \geq e{^2}+Q{^2}+a^2 \cos^2 \theta,
\end{eqnarray}
and both of the inner and outer ergosurfaces and the Cauchy and event horizons must obey the constraints of $r_{es-} \le r_{H-} \le r_{H+} \le r_{es+}$. In which, the outer boundary $r_{es+}$ is known as the static limit surface \cite{Griffiths}. In the case of extreme KNK BH, i.e., for the limiting value of spin parameter $a=\pm \sqrt {M^2-e^2-Q^2}$, the horizons of KNK BH coincides and is equal to its mass, while for $cos^2\theta=1$ (on rotational axis), $r_{es\pm}=r_{H\pm}$. On the equatorial plane $(\theta=\pi/2)$ of the KNK BH, the inner and outer ergosurfaces of the KNK BH modifies to
\begin{eqnarray}\label{sls1}
r_{es\pm}=M \pm \sqrt{ M^2-e{^2}-Q{^2}}.
\end{eqnarray}
\begin{figure*}
\begin{minipage}[b]{0.58\textwidth} \hspace{0.3cm}
\includegraphics[width=.75\textwidth]{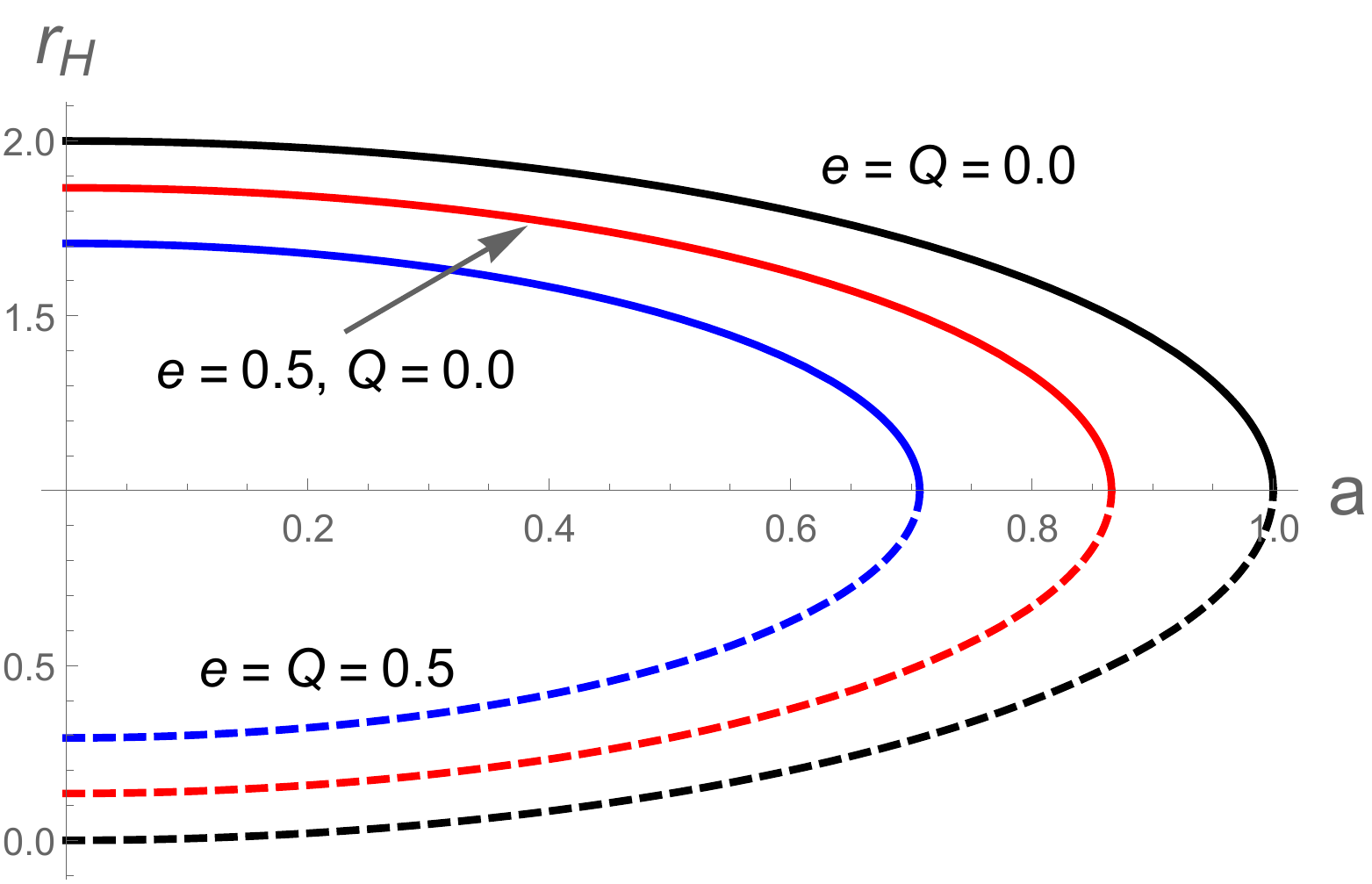}
\end{minipage}
\vspace{0.4cm}
\begin{minipage}[b]{0.58\textwidth} \hspace{-1.1cm}
\includegraphics[width=0.75\textwidth]{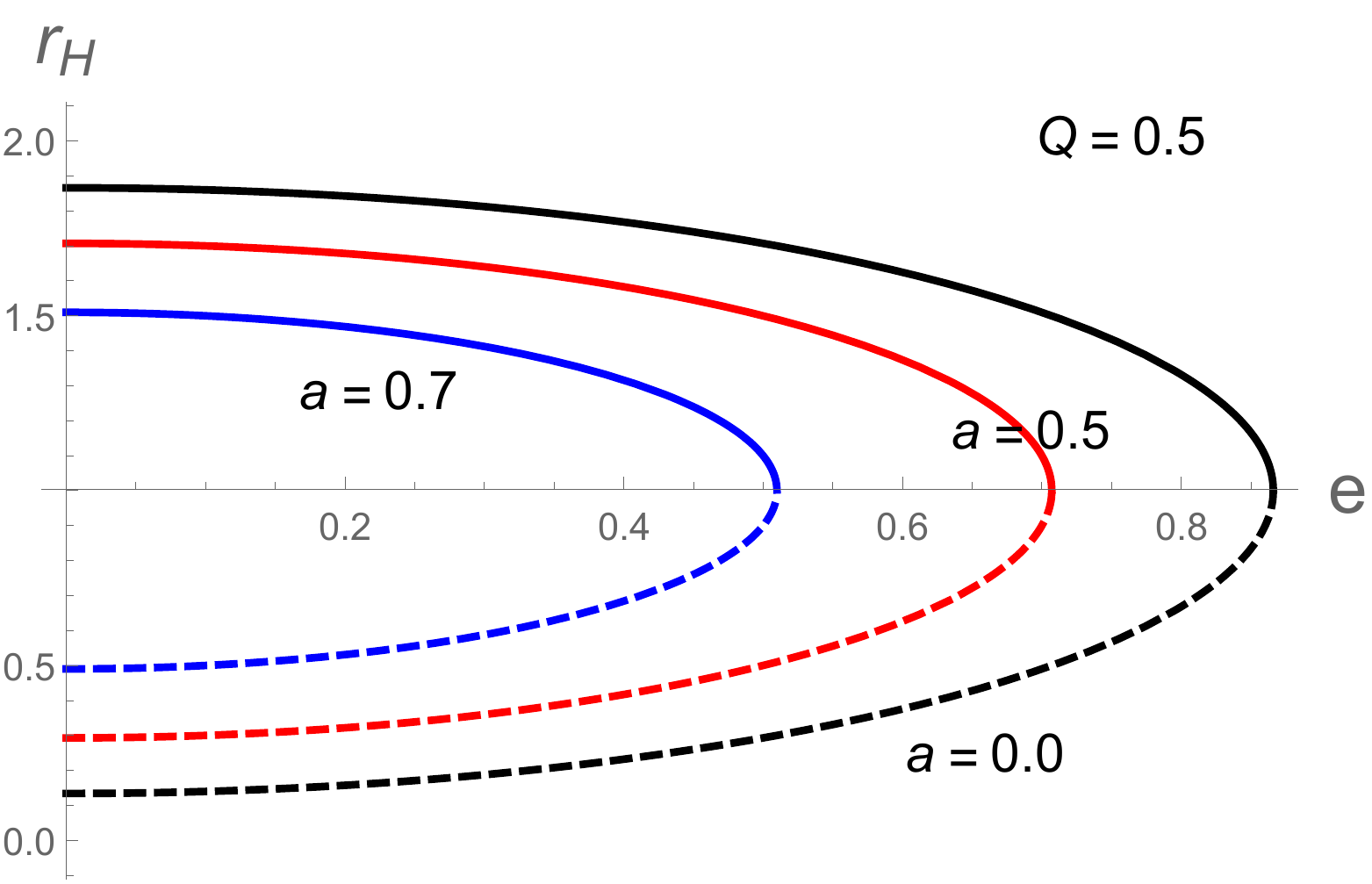}
\end{minipage}
\begin{minipage}[b]{0.58\textwidth} \hspace{0.3cm}
\includegraphics[width=.75\textwidth]{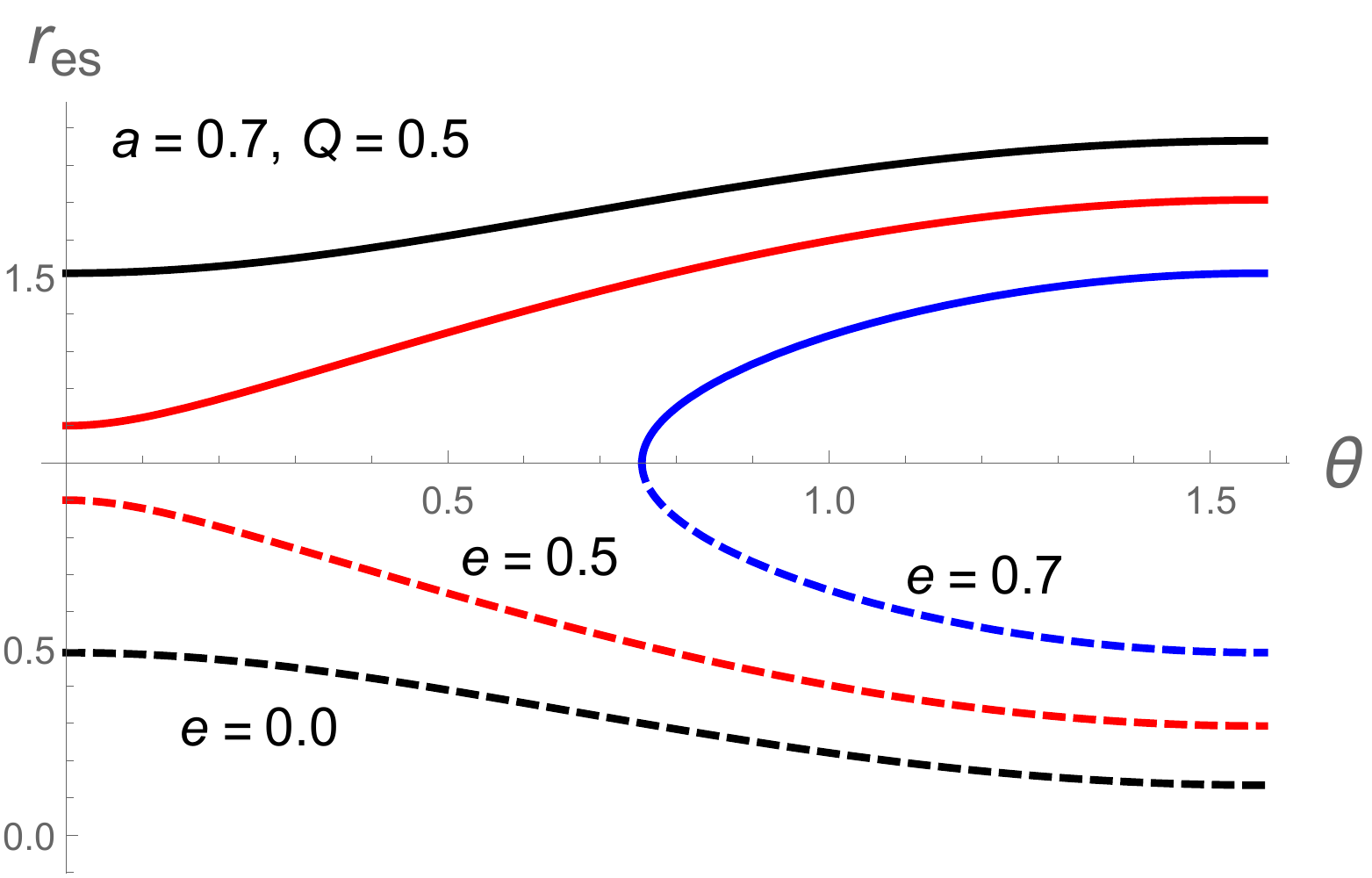}
\end{minipage}
\begin{minipage}[b]{0.58\textwidth} \hspace{-1.1cm}
\includegraphics[width=.75\textwidth]{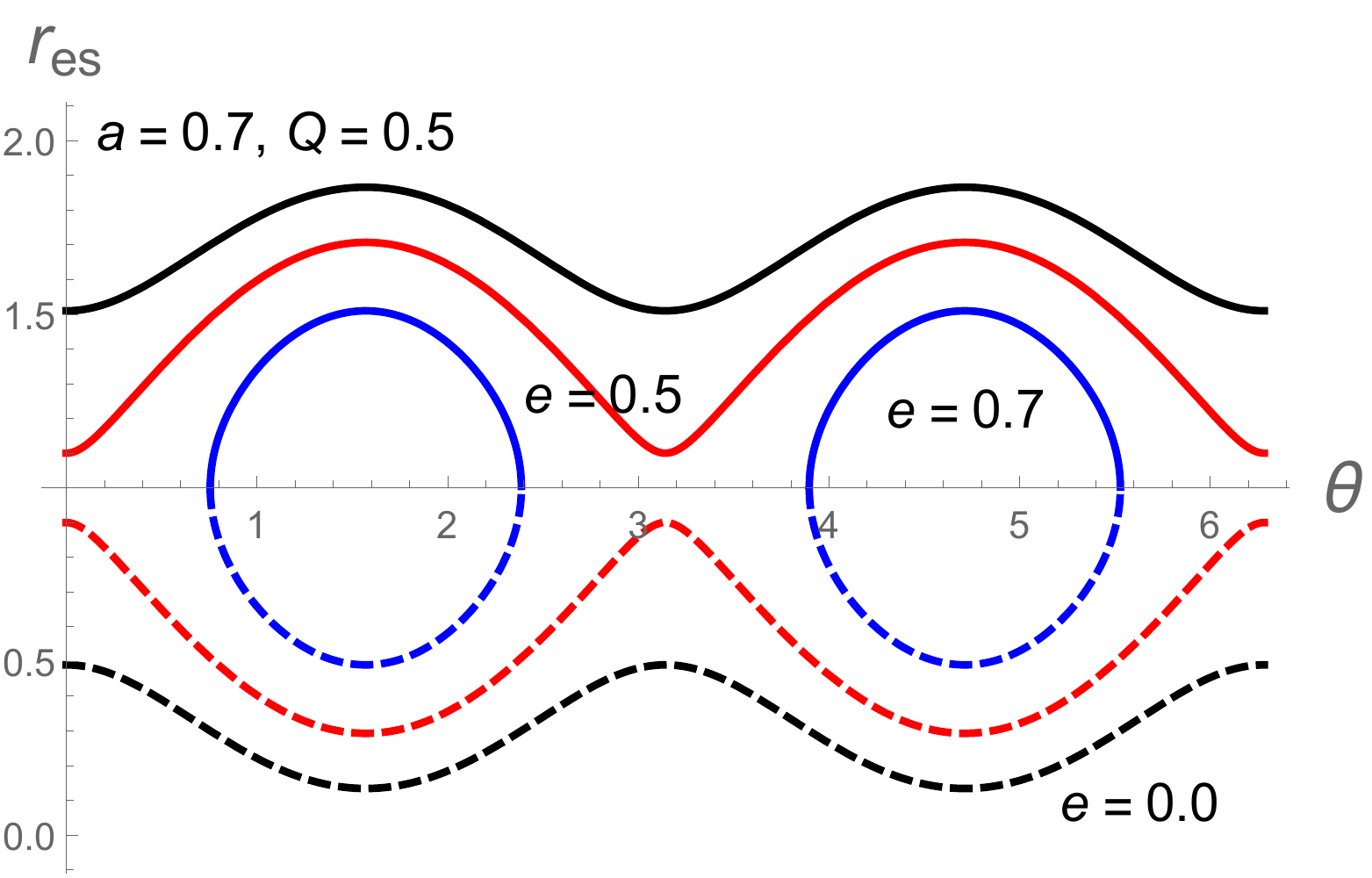}
\end{minipage}
\caption{Graphical description of the event and Cauchy horizons, (respectively the solid and dashed curves) upper row. The inner and outer ergosurfaces, (respectively the dashed and solid curves) lower row.}\label{Horizons2D}
\end{figure*}
From the above equation, it can be noted that on the equatorial plane the inner and outer ergosurfaces $r_{es\pm}$, are independent of BH rotation. In general, for rotating BHs the region $\delta\in (r_{H+},  r_{es+})$, is known as the ergoregion where no observer can remain static. This means that there could be no static observer with $d\theta/d\tau = d\phi/d\tau=dr/d\tau=0$ (with $\tau$ is the proper time).
On considering BH spacetime, the $r$-coordinate will be time-like if $r\in (r_{H-}, r_{H+})$, while space-like if  $r \in (0, r_{H-})\cup r > r_{H+} $.
This simply implies that the surfaces of constant $r$, i.e., $\delta_r$, is space-like for $\Delta < 0$, time-like for $\Delta > 0$ and light-like for $\Delta=0$ \cite{Pugl3}. Another characteristic of the ergoregion is that the test particle may have negative energy with respect to an observer at infinity.
Therefore, from the physical point of view, the ergoregion can allow the Penrose process of energy extraction from the KNK spacetime. Moreover, one must note that particles with positive energy can enter or leave the ergoregion, whereas particles with negative energy may not be able to leave the ergoregion. Particles will have negative energy within ergoregion, if and only if angular momentum $L$ of the particle is also negative. Specifically, on considering the circular geodesics within ergoregion both $\mathcal{E}$ and $L$ should be negative \cite{Chandrasekhar}.

\begin{figure*}
\begin{minipage}[b]{0.58\textwidth} \hspace{0.3cm}
\includegraphics[width=.75\textwidth]{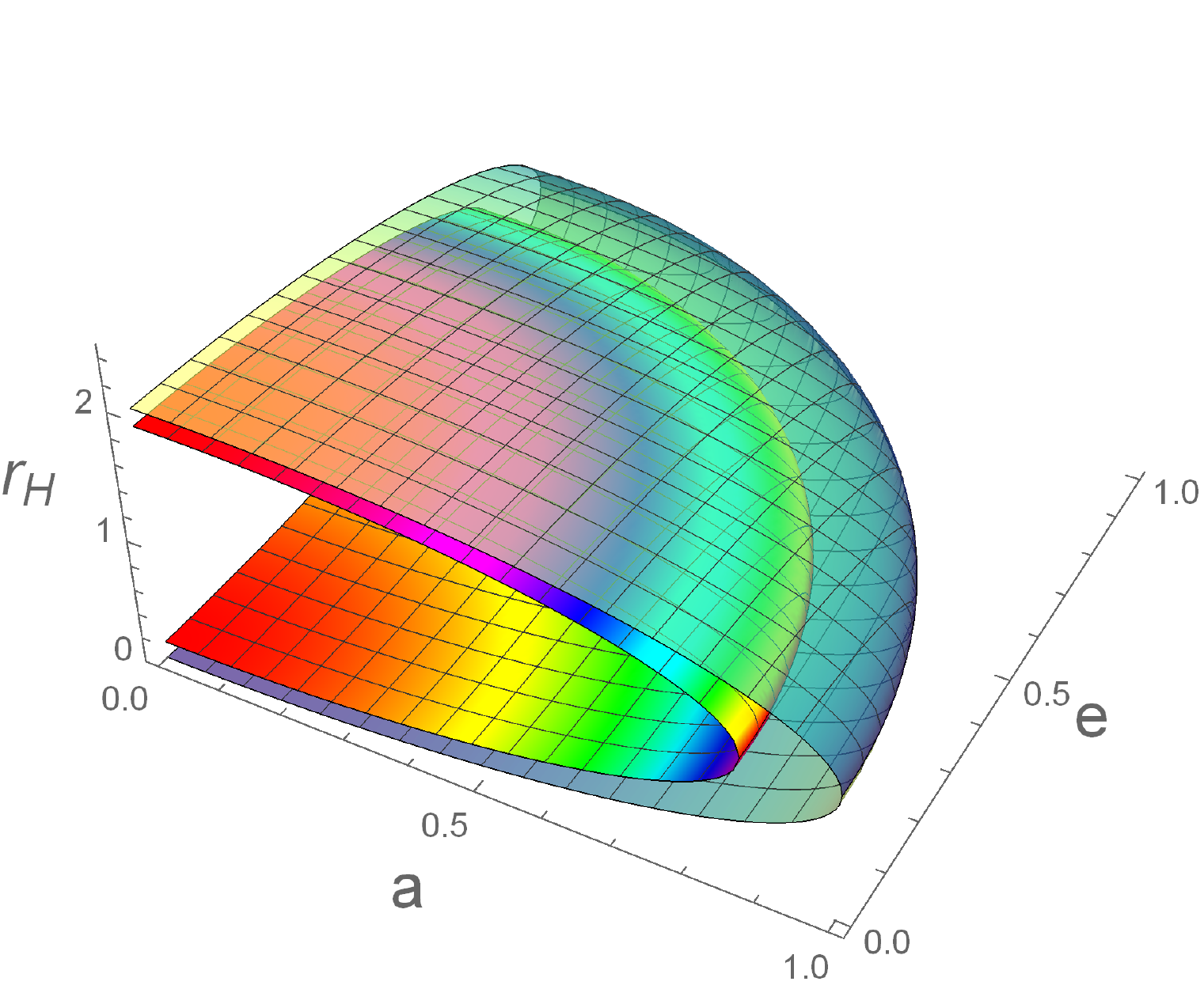}
\end{minipage}
\vspace{0.3cm}
\begin{minipage}[b]{0.58\textwidth} \hspace{-1.1cm}
\includegraphics[width=0.75\textwidth]{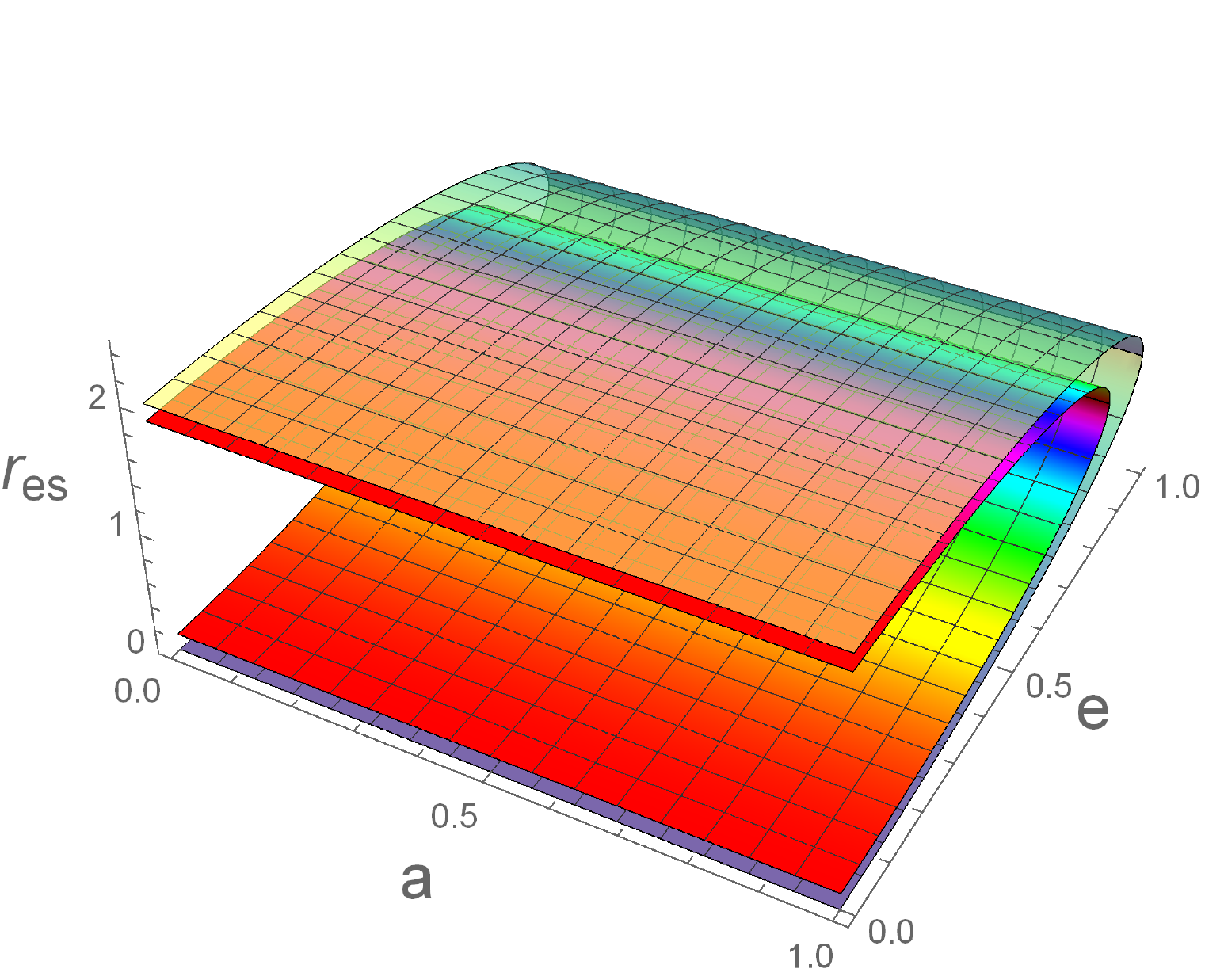}
\end{minipage}
\begin{minipage}[b]{0.58\textwidth} \hspace{0.3cm}
\includegraphics[width=.75\textwidth]{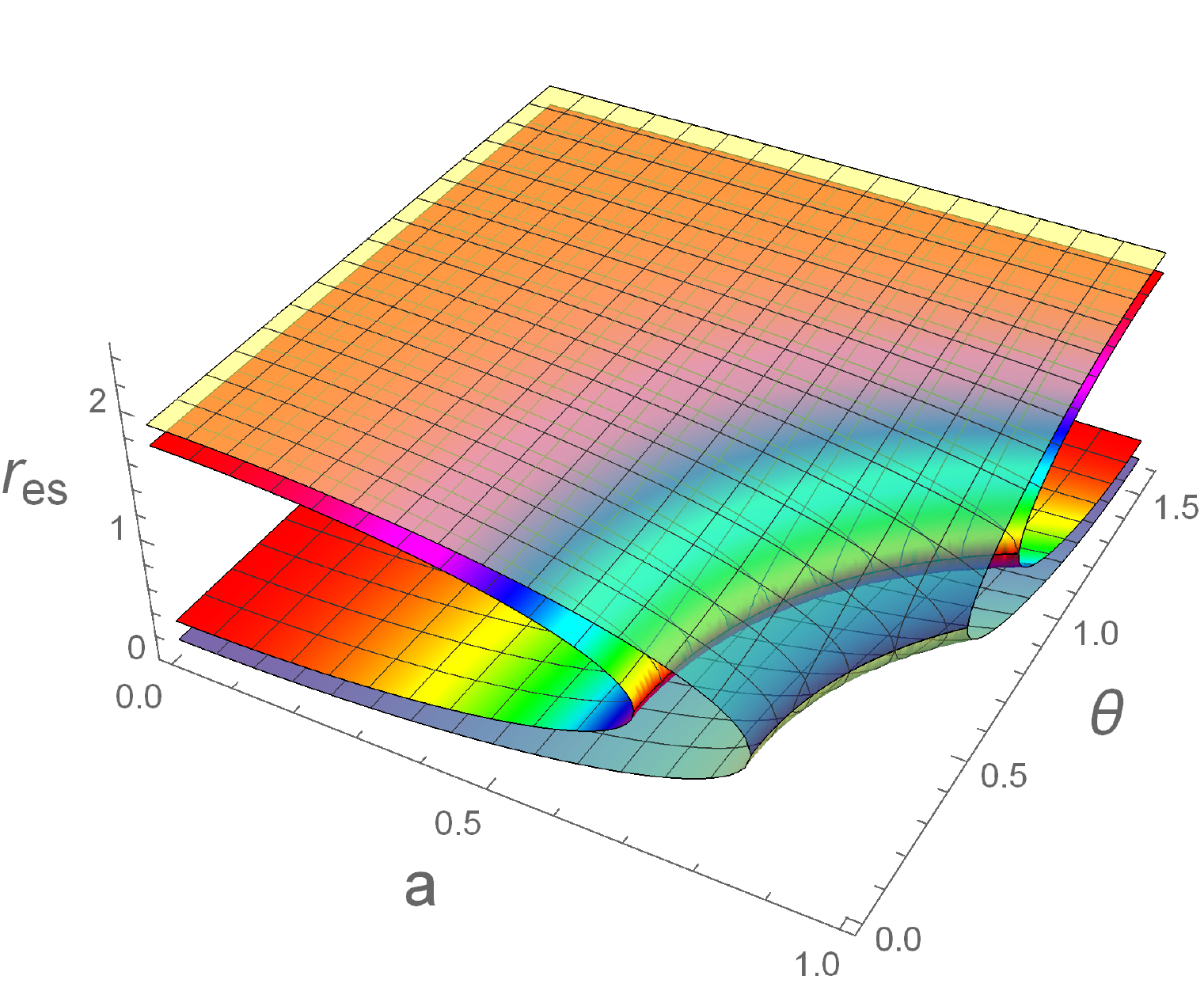}
\end{minipage}
\begin{minipage}[b]{0.58\textwidth} \hspace{-1.1cm}
\includegraphics[width=0.75\textwidth]{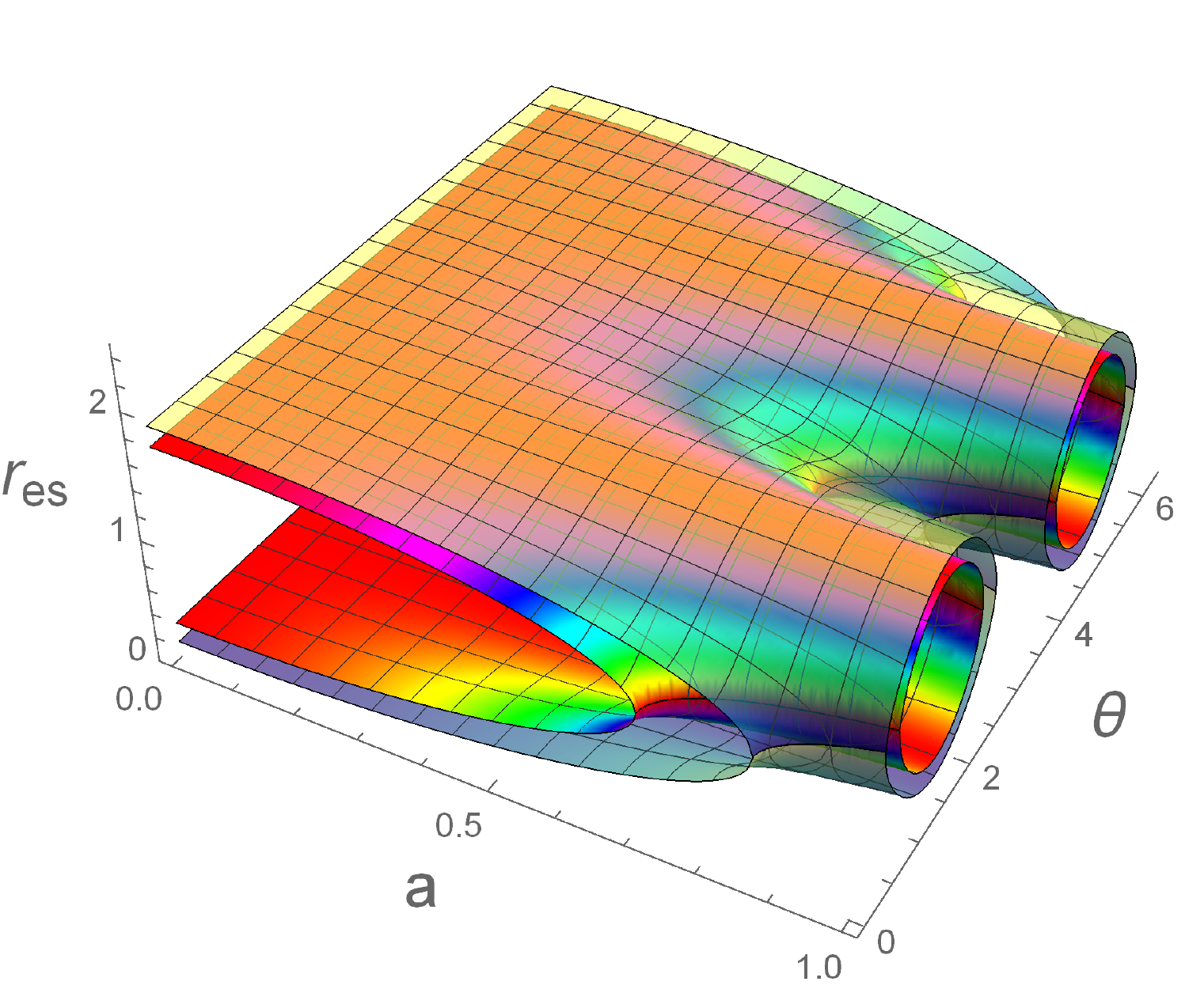}
\end{minipage}
\caption{3D view of the horizons $r_{H\pm}$ (upper left panel) and ergosurfaces $r_{es\pm}$, at $\theta=\pi/2$ (upper right panel). The inner and outer ergosurfaces, respectively the lower and upper surfaces (lower row), for different ranges of $\theta$. The outer and inner surfaces are, respectively plotted at $Q=0$ and $Q=0.5$.}\label{Horizons3D}
\end{figure*}

Figures \ref{Horizons2D} and \ref{Horizons3D}, respectively reflect the 2D and 3D representations of the inner and outer horizons, as well as the inner and outer ergosurfaces of the KNK spacetime. In Fig. \ref{Horizons2D}, it is depicted that BH rotation and dyonic charge, contributes to both the inner horizon and ergosurface (respectively shown in the upper and lower row) whereas, decreases its outer horizon and ergosurface (respectively shown in the upper and lower row). In Fig.  \ref{Horizons3D}, the upper and lower surfaces, respectively represent the outer and inner horizons (upper left channel), while the outer and inner ergosurfaces (remaining channels). We observed that BH rotation diminishing both width and diameter of the KNK BH horizons and ergosurfaces.
\section{Geodesics motion}
\label{sec:3}
The main purpose of this section is to study particle dynamics and the geodesics motion  around a KNK BH. Since they are of vital importance in both BH physics and in the theory of accretion disc to discover the essential attributes of a BH. The neutral particle motion can be described by the Lagrangian equation as
\begin{equation}\label{BS1}
\mathcal{L} = \frac{1}{2}g_{\mu\nu}\dot{x}^{\mu}\dot{x}^{\nu},
\end{equation}
here $\dot{x}^{\mu}={d x^{\mu}}/{d \tau}$ denotes the four-velocity with dot means $\partial / \partial \tau$. For simplicity, we fix our attention to the equatorial plane. Since both $t$ and $\phi$ are cyclic coordinates with respect to the constants of motion, i.e., energy $\mathcal{E}$ and angular momentum $L$, remain conserved along geodesics. Consequently, the corresponding generalized momenta take the form
\begin{eqnarray}\label{BS2}
\frac{\partial\mathcal{L}}{\partial\dot{t}}&=&p_{t}=g_{tt}\dot{t}+g_{t\phi} \dot{\phi}=-\mathcal{E},\\ \label{BS3}
\frac{\partial\mathcal{L}}{\partial\dot{\phi}}&=&p_{\phi}=g_{t\phi}\dot{t}+g_{\phi\phi}\dot{\phi}=L,
\end{eqnarray}
On solving the above equations for $\dot{t}$ and $\dot{\phi}$, leads us to the following two equations
\begin{eqnarray}\label{BS4}
r^2\frac{dt}{d\tau}&=& a(L-a \mathcal{E})+\frac{(a^2+r^2)}{\Delta}\left((a^2+r^2)\mathcal{E}-aL \right),\\\label{BS5}
r^2\frac{d\phi}{d\tau}&=&  \left(L-a \mathcal{E} \right)+\frac{a}{\Delta}\left((a^2+r^2)\mathcal{E}-aL \right).
\end{eqnarray}
Using the normalization condition $u^\mu u_\mu=-m_0$ ($m_0$ is the particle's rest mass), it follows that
\begin{eqnarray}\label{BS8}
r^2\frac{d r}{d \tau}&=& \pm \sqrt{\mathcal{R}(r)},
\end{eqnarray}
with
\begin{eqnarray}\label{BS10}
\mathcal{R}(r)&=&\left((a^2+r^2)\mathcal{E}-a L\right)^2-\Delta \left(m_0^2r^2+(L-a\mathcal{E})^2 \right).
\end{eqnarray}  
The above three equations, i.e., Eqs. \eqref{BS4}, \eqref{BS5} and \eqref{BS8}, governing different types of geodesics equations.
For circular orbits, the following constraints must hold simultaneously
\begin{equation}\label{BS11}
\mathcal{R}(r)=0, \quad \partial_r \mathcal{R}(r)=0.
\end{equation}
Henceforth, using the above constraints and spacetime line element of the KNK BH. The specific energy $\mathcal{E}$ and angular momentum $L$, related to a distant observer can be obtained as \cite{Aliev, Stuch4}
\begin{eqnarray}\label{BS12}
&&\mathcal{E}=\frac{\Delta-a^2\pm a\mathcal{Z}}{r\sqrt{r^2-3Mr+2(e^2+Q^2)\pm 2a\mathcal{Z}}},\\\label{BS13}
&&L=\pm\frac{\mathcal{Z} \left(a^2+r^2\mp 2a\mathcal{Z}\right)\mp a(e^2+Q^2)}{r\sqrt{r^2-3Mr+2(e^2+Q^2)\pm 2a\mathcal{Z}}}.
\end{eqnarray}
In the above equations $\mathcal{Z}=\sqrt{Mr-e^2-Q^2}$, while the upper and lower signs, at a large distance, respectively correspond to the co-rotating and counter-rotating orbits.
It can be observed from Eqs. \eqref{BS12} and \eqref{BS13}, that the existence of circular geodesics requires the restrictions:
\begin{eqnarray}\label{BS14}
&&{r^2-3Mr+2(e^2+Q^2)\pm 2a\mathcal{Z}} \geq 0,\\\label{BS15}
&&Mr \geq {e^2+Q^2}.
\end{eqnarray}
In the first equation, equality governs the circular geodesics of photons, whereas the second restriction is related to the region of positive radii. In the case of KNK spacetime equality of the second condition locates the innermost stable circular orbits.
\subsection{Photon and marginally stable circular orbits}
\begin{figure*}
\begin{minipage}[b]{0.58\textwidth} \hspace{0.3cm}
        \includegraphics[width=0.75\textwidth]{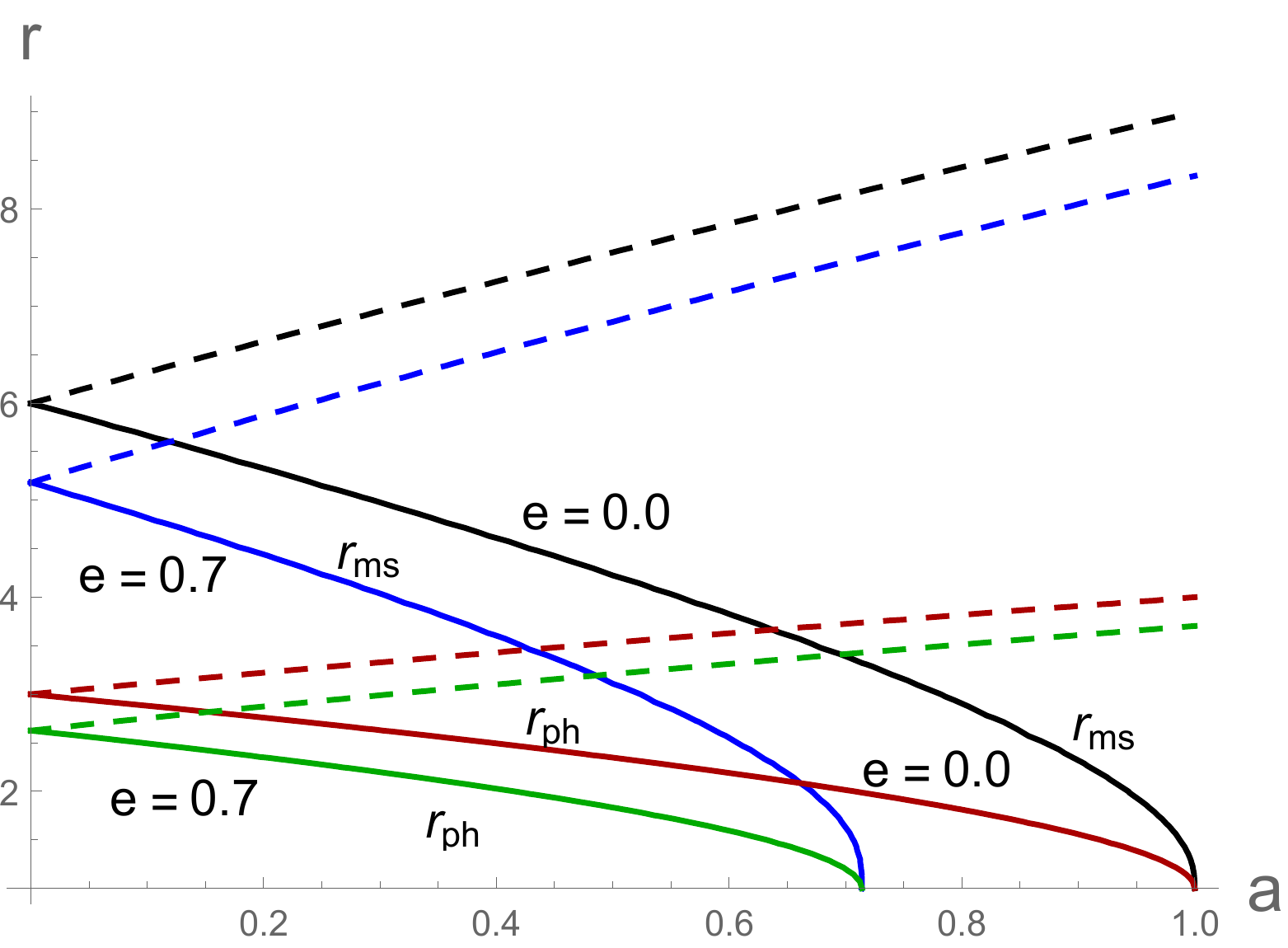}
    \end{minipage}
    \vspace{0.5cm}
        \begin{minipage}[b]{0.58\textwidth} \hspace{-1.1cm}
       \includegraphics[width=.75\textwidth]{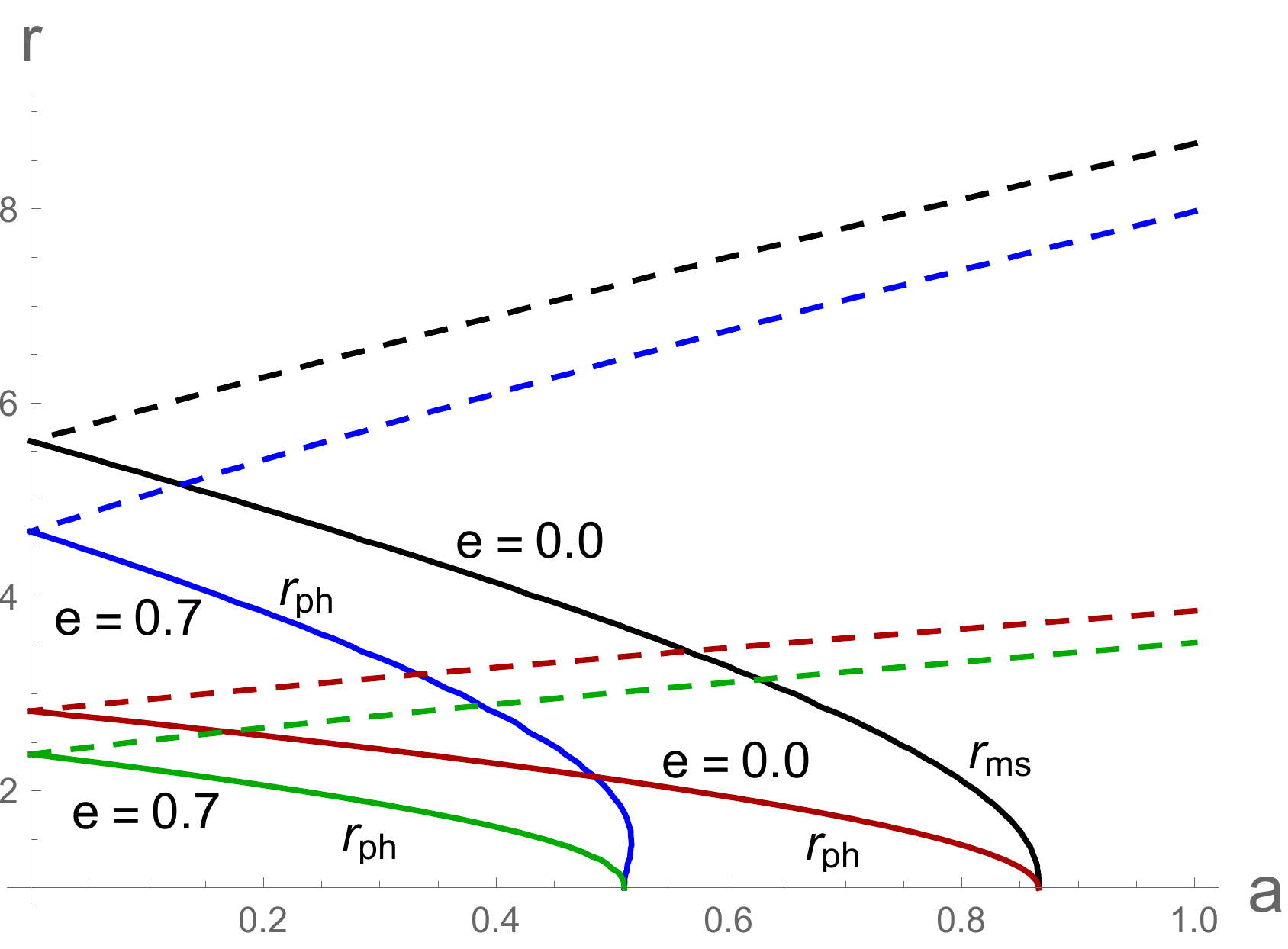}
    \end{minipage}
\caption{Plots showing the comparison of $r_{ph}$ with $r_{ms}$, both for prograde orbits (solid curves) and retrograde orbits (dashed curves) along $a$. The left channel is for KN BH at $Q=0$, while the right one is for KNK BH at $Q=0.5$.}\label{rr}
\end{figure*}

On the equatorial plane, photon orbits can be obtained from the radial function $\mathcal{R}(r)$, as given by Eq. \eqref{BS10}, with the assumption of $m_0=0$ (photon mass). As a result, radial Eq. \eqref{BS10} can be rewritten as
\begin{eqnarray}\label{Ph1}
\frac{\mathcal{R}(r)}{\mathcal{E}^2}=\left((a^2+r^2)-a \ell\right)^2- \Delta(\ell-a)^2,
\end{eqnarray}
where $\ell$ is the impact parameter defined by $\ell=L/\mathcal{E}$. Moreover, it is important to note the direct dependency of photon orbits on $\ell$. The equatorial circular photon orbits can be obtained by imposing the conditions of circular motion defined in Eq. \eqref{BS14}, as a result
\begin{eqnarray}\label{Ph2}
\left(a^2+r^2-a\ell\right)^2-\Delta(\ell-a)^2=0,\\\label{Ph3}
2r\left(a^2+r^2-a\ell\right)-(r-M)(\ell-a)^2=0.
\end{eqnarray}
By following \cite{Sche,Blaschke}, one can obtain the circular geodesics of photon around a KNK BH:
\begin{equation}\label{Ph3}
r^2-3Mr+2(e^2+Q^2)\pm 2a\sqrt{Mr-e^2-Q^2}= 0.
\end{equation}
The above equation requires the same reality condition of circular photon orbits $r_{ph}$, like the one that follows from Eqs. \eqref{BS12} and \eqref{BS13}, i.e., $r_{ph}\geq (e^2+Q^2)/M$.
On solving Eq. \eqref{Ph3} for the spin parameter $a$, of the BH leads us to
\begin{equation}\label{Ph4}
a=a_{ph}=\pm \frac{3Mr-r^2-2(e^2+Q^2)}{2\sqrt{Mr-e^2-Q^2}}.
\end{equation}
For $a>0$, the above function determines the radii of both prograde and retrograde circular photon orbits. Local extrema of $a_{ph}$, can be obtained at $r=1$ and $r=4(e^2+Q^2)/3$, while
the zeros of Eq. \eqref{Ph4} can be obtained as
\begin{equation}\label{Ph5}
r_{ph\pm}=\frac{1}{2}(3M\pm \sqrt{9M^2-8(e^2+Q^2)}).
\end{equation}
The circular stable orbits of geodesics motion governed by the constraints
\begin{equation}\label{Ph6}
\frac{\partial^2 \mathcal{R}(r)}{\partial r^2} \leq 0,
\end{equation}
in which equality determines the marginally stable circular orbits (MSCO). Henceforth, by making use of condition $\partial^2 \mathcal{R}(r) / \partial r^2=0$, the relation of MSCO for a KNK BH can be obtained as
\begin{eqnarray}\label{Ph7}
\left(3a^2+6Mr-r^2-9(e^2+Q^2)\right)Mr-4(a^2-e^2-Q^2)(e^2+Q^2) \mp 8a (Mr-e^2-Q^2)^{3/2}=0.
\end{eqnarray}
From the above equation, we get
\begin{equation}\label{Ph8}
a=a_{ms}=\mp \frac{4(Mr-e^2-Q^2)^{3/2}\pm\sqrt{\mathcal{D}}}{4(e^2+Q^2)-3Mr},
\end{equation}
in which
\begin{equation*}
\mathcal{D}=3M^2r^2(e^2+Q^2)-(2M^3+4M(e^2+Q^2))r^3+3Mr^4.
\end{equation*}
While the upper and lower family of geodesics are represented by $\mp$ sings, whereas the possible solutions of Eq. \eqref{Ph7} are represented by the $\pm$ signs. By considering a KN braneworld BH, the circular equatorial orbits are numerically analyzed and concluded that in the case of positive brane parameter, the radii of both photon orbits and the MSCOs moves towards the event horizon \cite{Aliev}. We have observed the same behaviour for the KNK BH, with the increasing values of electric charge, while more rapid fall down of orbits towards the event horizon in the presence of magnetic charge $Q$. In the case of Schwarzschild BH, for both prograde and retrograde orbits, we get  $r_{ph}=3M$ and $r_{ms}=6M$. 
\begin{figure*}
\begin{minipage}[b]{0.58\textwidth} \hspace{0.3cm}
        \includegraphics[width=0.75\textwidth]{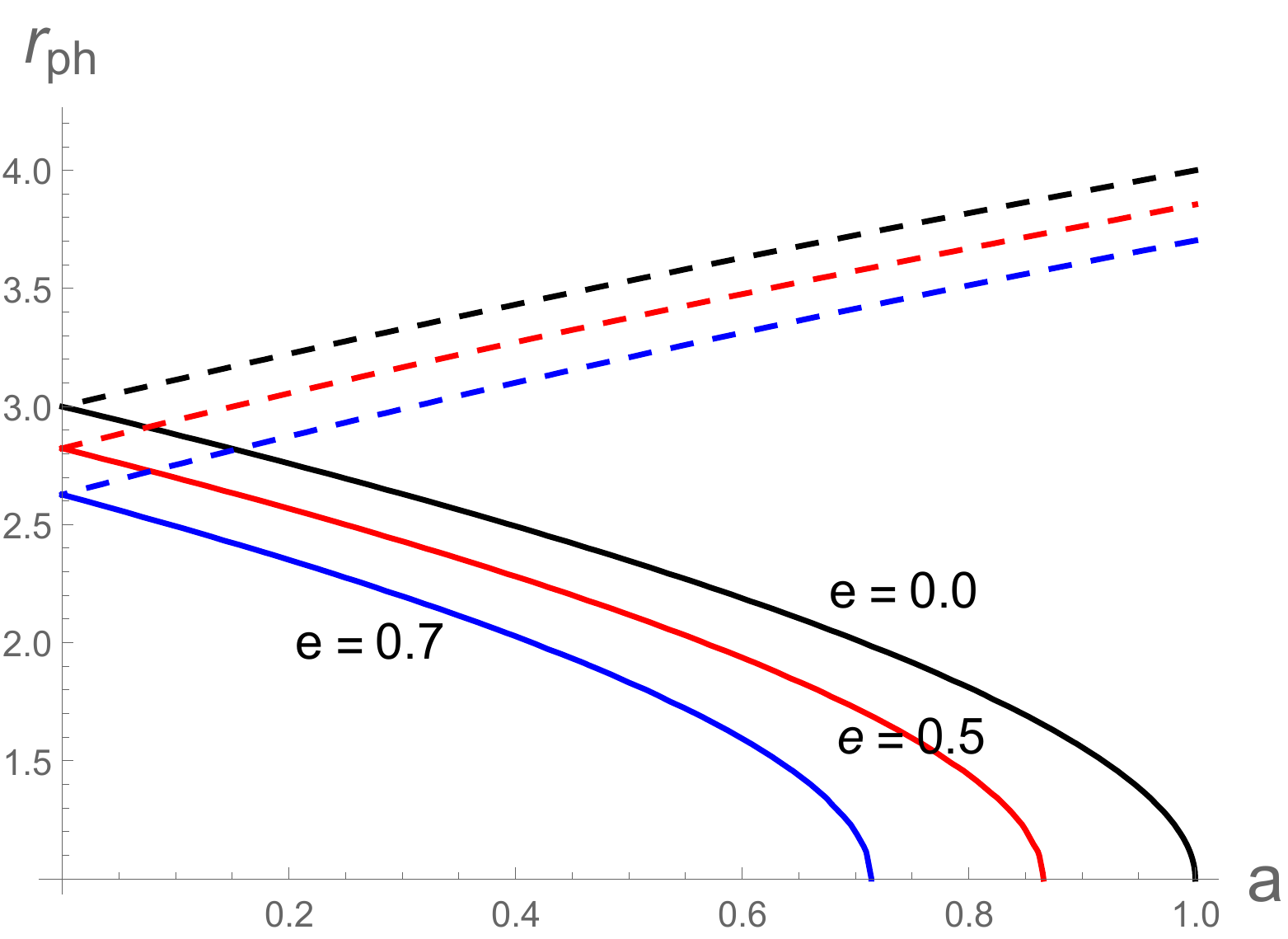}
    \end{minipage}
    \vspace{0.5cm}
        \begin{minipage}[b]{0.58\textwidth} \hspace{-1.1cm}
       \includegraphics[width=.75\textwidth]{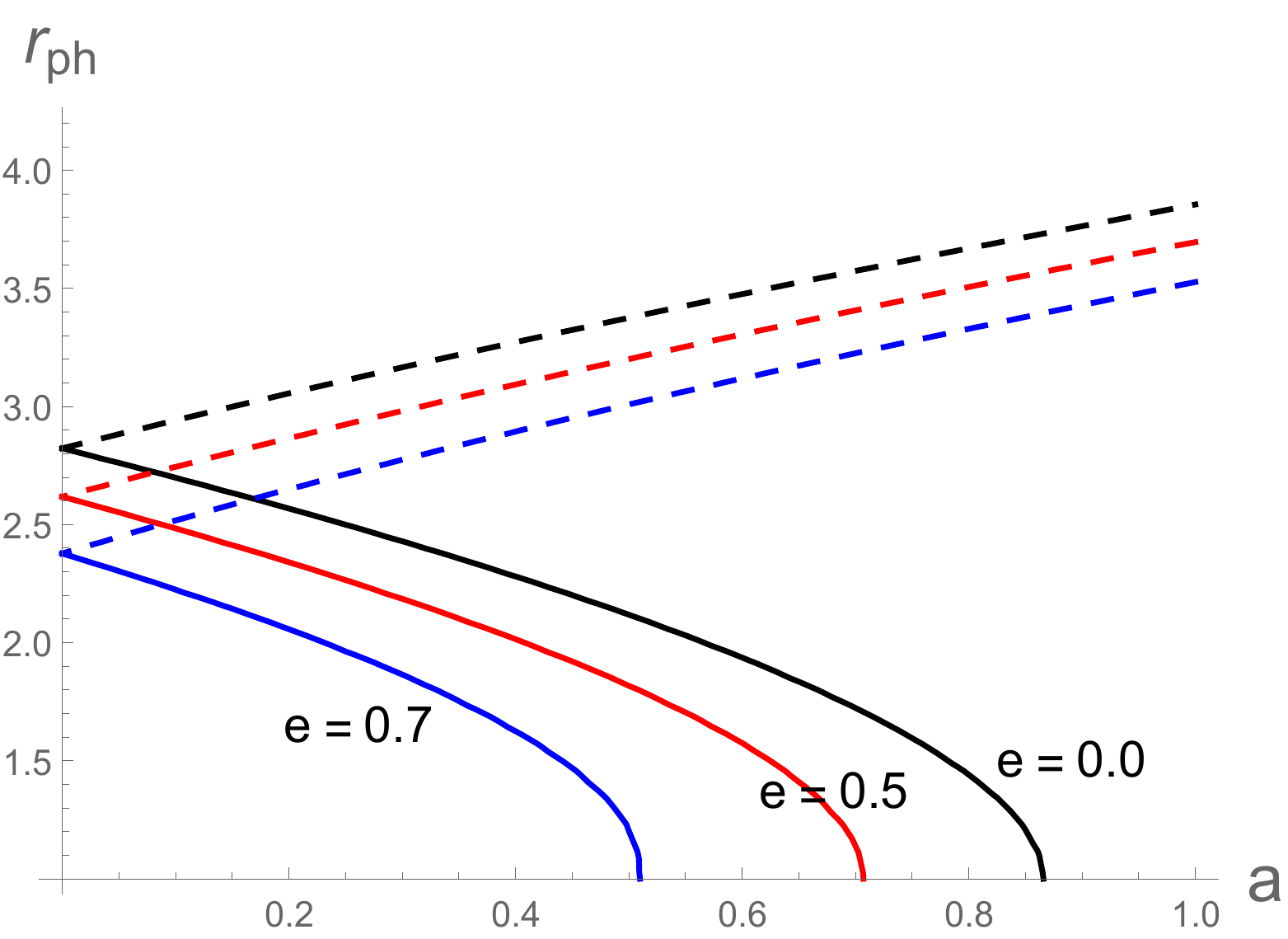}
    \end{minipage}
\begin{minipage}[b]{0.58\textwidth} \hspace{0.3cm}
        \includegraphics[width=0.75\textwidth]{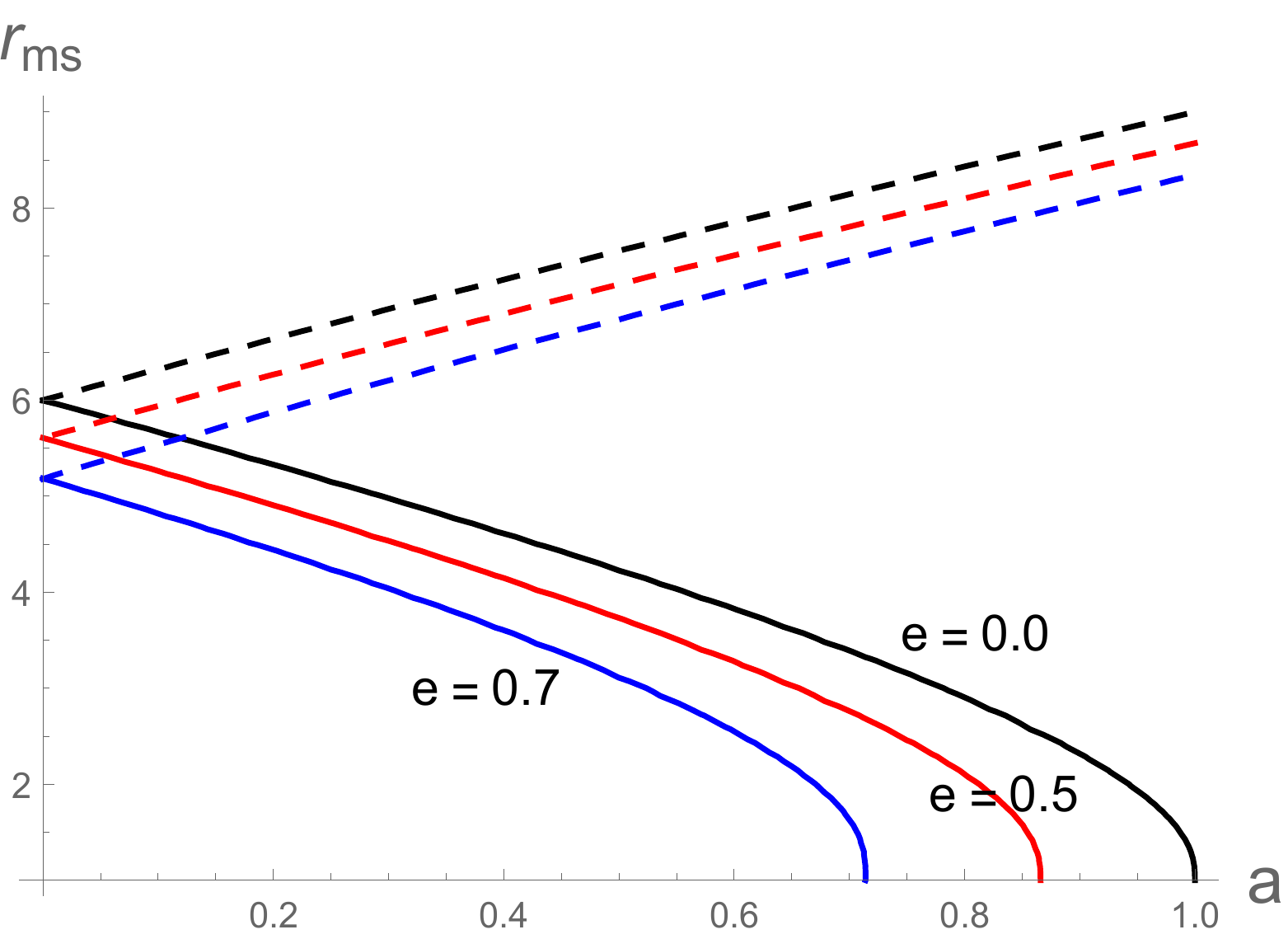}
    \end{minipage}
        \begin{minipage}[b]{0.58\textwidth} \hspace{-1.1cm}
       \includegraphics[width=.75\textwidth]{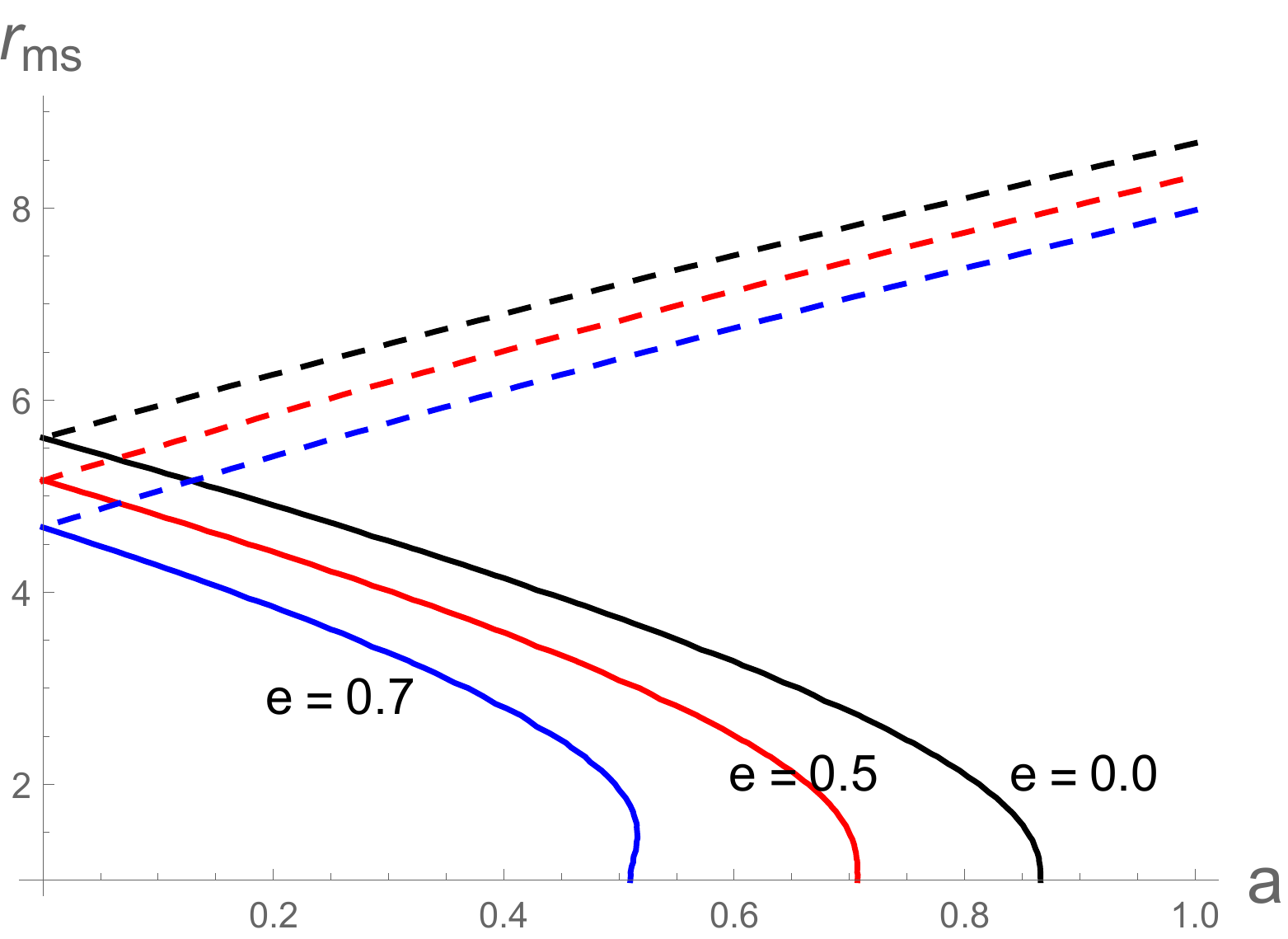}
    \end{minipage}
\caption{Plots showing the behaviour of $r_{ph}$ and $r_{ms}$, both for prograde orbits (solid curves) and retrograde orbits (dashed curves) along $a$. The left channels are plotted for KN BH at $Q=0$, while the right channels are for KNK BH at $Q=0.5$.}\label{rph}
\end{figure*}
The circular photon orbit $r_{ph}$ and MSCO $r_{ms}$, are plotted in Fig. \ref{rph}, as a function of BH $a$, at different values of the dyonic charge parameters. In the left channels, the black curves correspond to the Kerr BH ($e=Q=0$), while in the right channels they correspond to the KN BH ($e\neq0, Q=0$). It can be observed that BH charge diminishing the radii of photon orbits, as well as the MSCOs of both prograde and retrograde particle motions. On the other hand, BH spin diminishing the radii of both photon orbits and the MSCOs in the case of direct particle motions, while contributes to the radii of retrograde particle motions. Moreover, it can be observed that the KNK BH have smaller radii of the photon orbits and the MSCOs, as compared to the KN and Kerr BHs. A comparative analysis of the photon orbits with the MSCOs is given in Fig. \ref{rr}, for both prograde and retrograde motion of particles.
\subsection{Effective potential}
The effective potential (${V}_{eff}$) of a particle has important consequences in the physics of BH spacetimes. Since one can utilize the techniques of effective potential to study the characteristic of circular orbits and the range of its angular momentum. Henceforth, the effective potential of a particle can be  defined as \cite{Misner}
\begin{equation}\nonumber
\left(\frac{d r}{d \tau}\right)^{2}+{V}_{eff}=\mathcal{E}^2.
\end{equation}
By making use of Eq. \eqref{BS8}, ${V}_{eff}$ takes the form
\begin{equation}
{V}_{eff}=\frac{\mathcal{E}^2r^4+\Delta\left(L-a\mathcal{E}\right)^2- \left(\left(a^2+r^2\right)\mathcal{E}-aL\right)^2}{r^4}.
\end{equation}
\begin{figure*}
\begin{minipage}[b]{0.58\textwidth} \hspace{0.3cm}
        \includegraphics[width=0.75\textwidth]{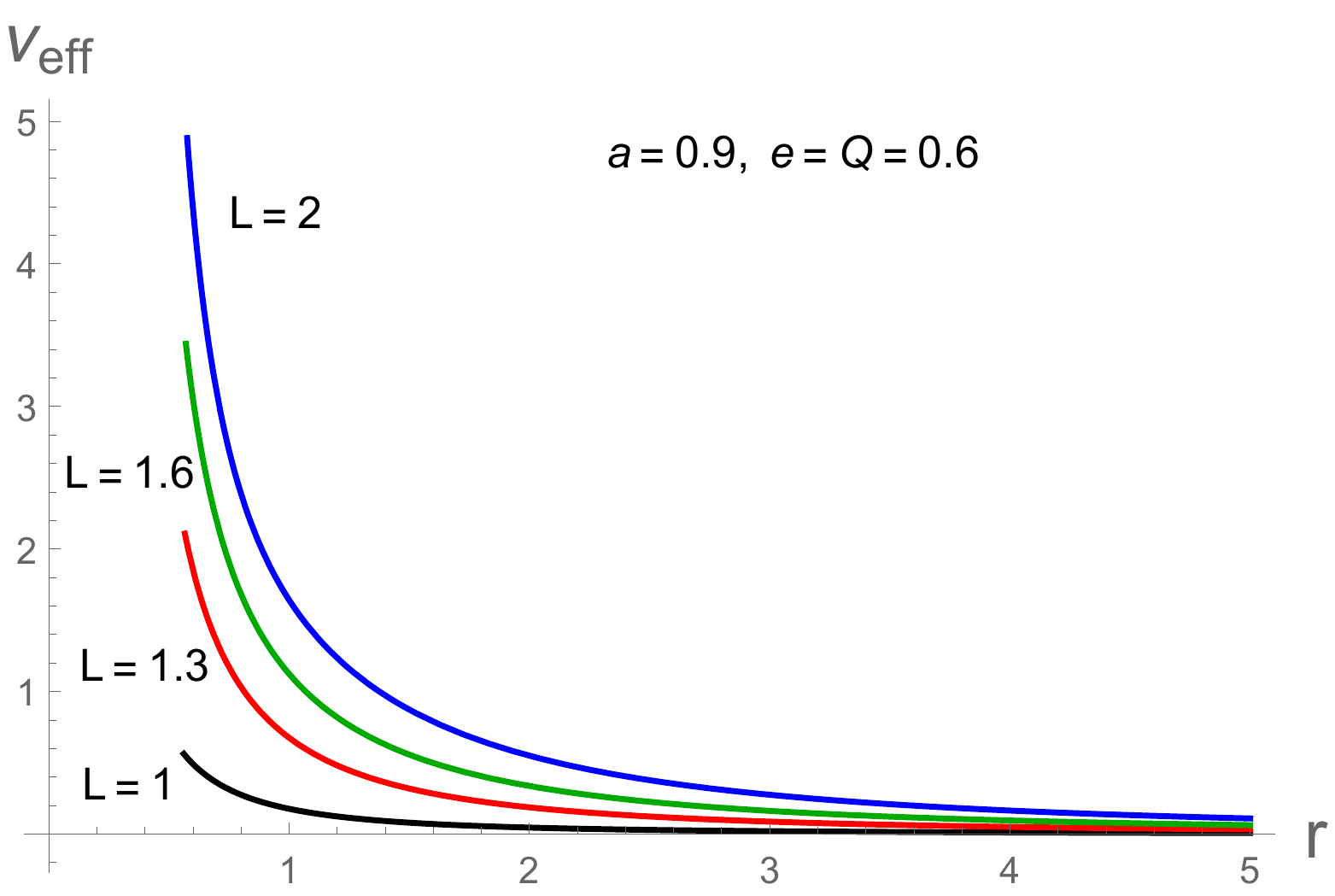}
    \end{minipage}
    \vspace{0.5cm}
        \begin{minipage}[b]{0.58\textwidth} \hspace{-1.1cm}
       \includegraphics[width=.75\textwidth]{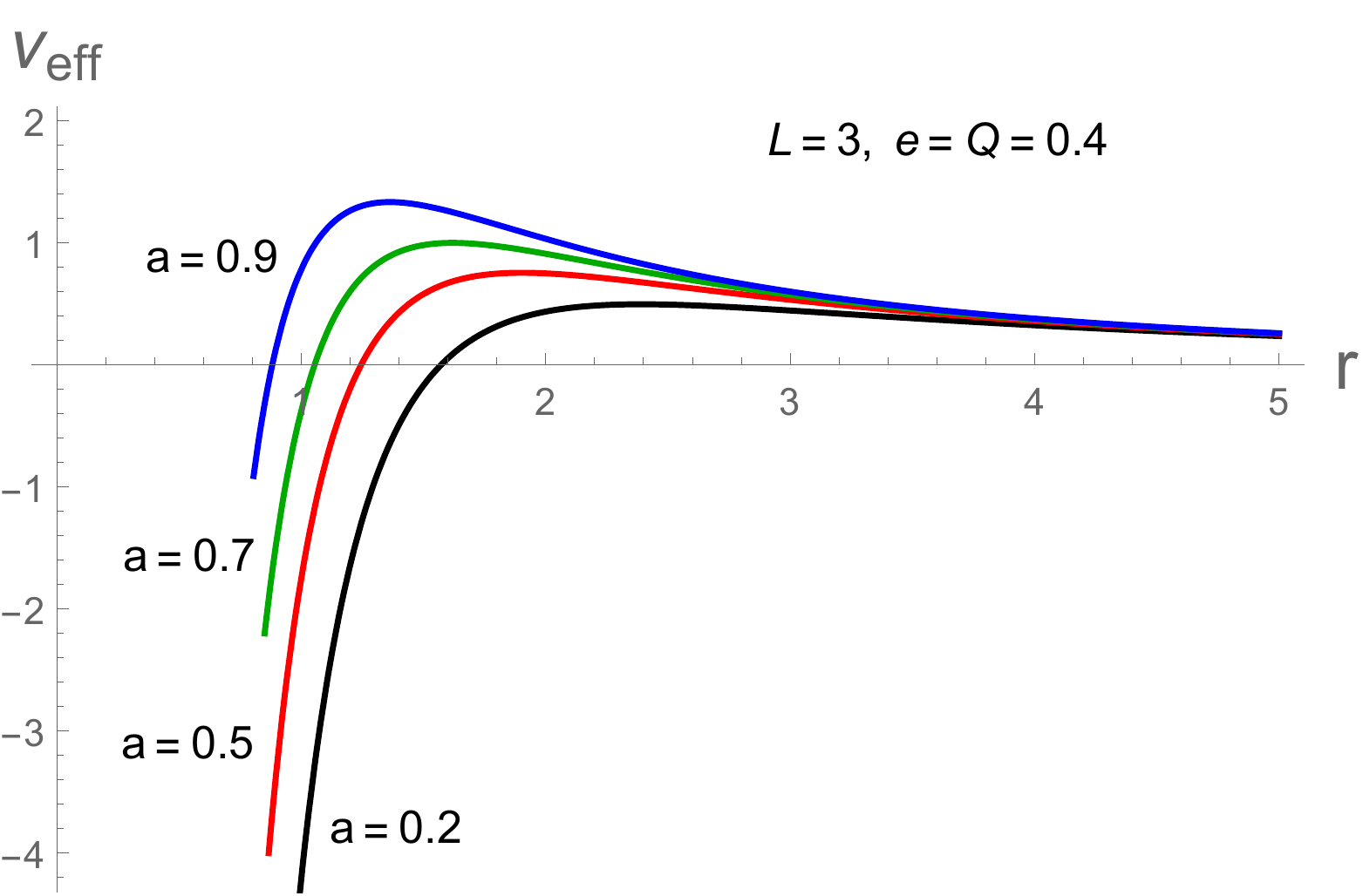}
    \end{minipage}
\begin{minipage}[b]{0.58\textwidth} \hspace{0.3cm}
        \includegraphics[width=0.75\textwidth]{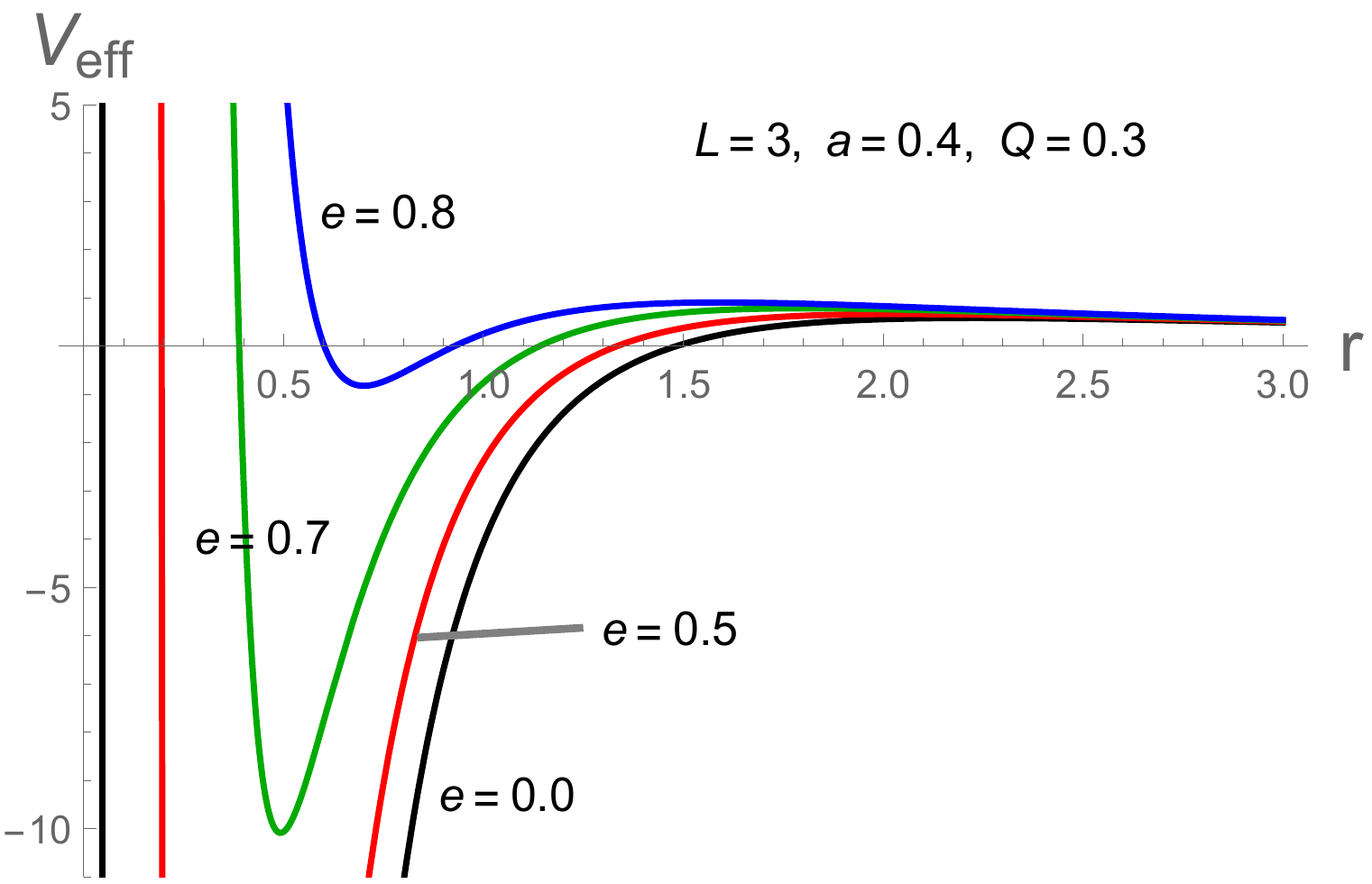}
    \end{minipage}
        \begin{minipage}[b]{0.58\textwidth} \hspace{-1.1cm}
       \includegraphics[width=.7\textwidth]{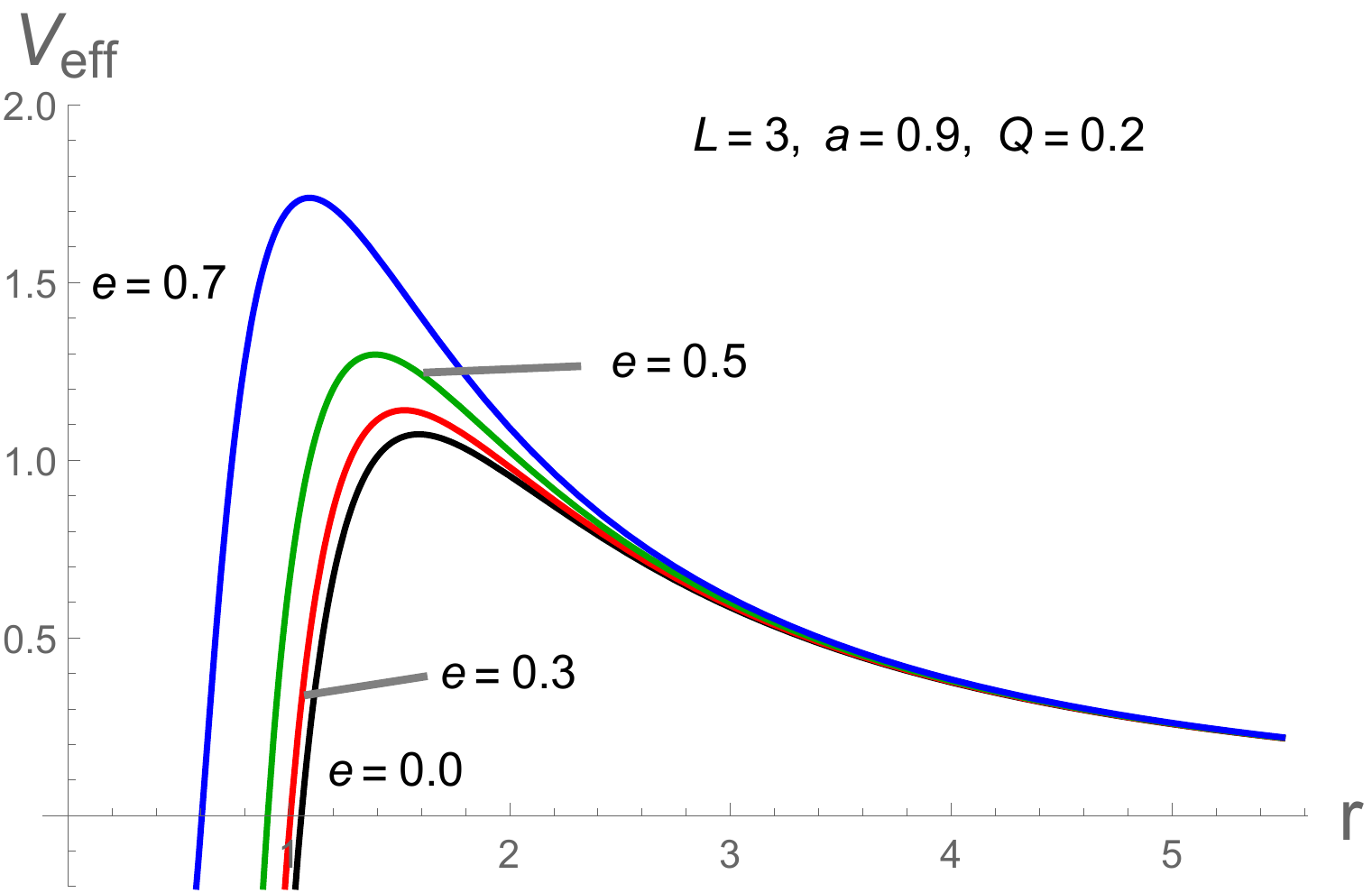}
    \end{minipage}
\caption{Plots showing the behaviour of ${V}_{eff}$ along radial distance $r$, with the assumption of $\mathcal{E}=1$.}\label{NEP}
\end{figure*}
The radial dependence of revolving particles effective potential around a KNK BH, at various values of spin and dyonic charge parameters are described in Fig. \ref{NEP}. From Eq. \ref{BS12}, it is clear that $\mathcal{E}$ can take different fixed values, so for simplicity, we assume $\mathcal{E}=1$ \cite{Toshmatov}. In the upper left panel, the effective potential is plotted for different values of the angular momentum $L$. For the variation of $a$, the curves coincide near $r = 3.5$ (upper right panel). The curves leading to the unstable orbits with their maximum values ${V}_{eff}\approx 1.33, \, 1.0,\, 0.75,\, 0.49$, corresponding to $a=0.9,\, 0.7,\, 0.5,\, 0.2$. We figured out that in the unstable case, BH rotation contributes to the instability of effective potential.
In the bottom channel, the effective potential is plotted for the variation of electric charge $e$ and its stability decrease with the increase of $e$ (left panel), whereas in the unstable case its maximum values are ${V}_{eff} \approx 1.73,\, 1.3,\, 1.14, \, 1.0$ (right panel). The behaviour of circular stable orbits in the case of null geodesics is also studied \cite{Sharif3}.  It is shown that in the unstable case its instability increase with $e$ and the curves coincides near $r=2.25$. It is observed that in the case of KNK BH, increasing the value of $e$, result in decreasing the stability, while contributes to the instability  of ${V}_{eff}$. In the lower channel, the black curves correspond to the effective potential of KN BH ($e=0$), showing more stable circular orbits as compared to the KNK BH.
\subsection{Lyapunov exponent}
The Lyapunov exponent can be used to study the average rate at which the nearby geodesics diverge or converge in the phase space. It illustrates the exponential rate of a trajectory's perturbation decays or grows at a certain point with time in the state space. The positive and negative Lyapunov exponents, respectively indicate divergence and convergence between nearby trajectories. Henceforth, using the equation of motion, the Lyapunov exponent of null geodesics can be obtained as \cite{Cardoso}
\begin{eqnarray}\nonumber
\lambda &=&\sqrt{\frac{-{V}_{eff}}{2 \dot{t}^2}}\\\label{Lexp}
&=&\frac{1}{L^2 r^6}[(\Delta-a^2)((10(e^2+Q^2)-12Mr)(L-a\mathcal{E})^2 +3r^2(L^2-a^2\mathcal{E}^2))]^{1/2}.
\end{eqnarray}
Since in the above equation, $(L-a\mathcal{E})^2 \geq 0$ and if simultaneously all the below three expressions are positive or any two of them are negative
\begin{equation}\nonumber
(r^2-2Mr+e^2+Q^2),\,\,\, (5(e^2+Q^2)-6Mr),\,\,\, (L^2-a^2\mathcal{E}^2),
\end{equation}
then $\lambda$ will be positive and as a result, the resultant circular orbits will turn out to be unstable. The Lyapunov exponent in Eq. \eqref{Lexp}, reduces into the KN BH by putting $Q=0$; to the Kerr BH if $e=Q=0$; and to the Schwarzschild BH if $a=e=Q=0$.
\begin{figure*}
\begin{minipage}[b]{0.58\textwidth} \hspace{0.3cm}
        \includegraphics[width=0.7\textwidth]{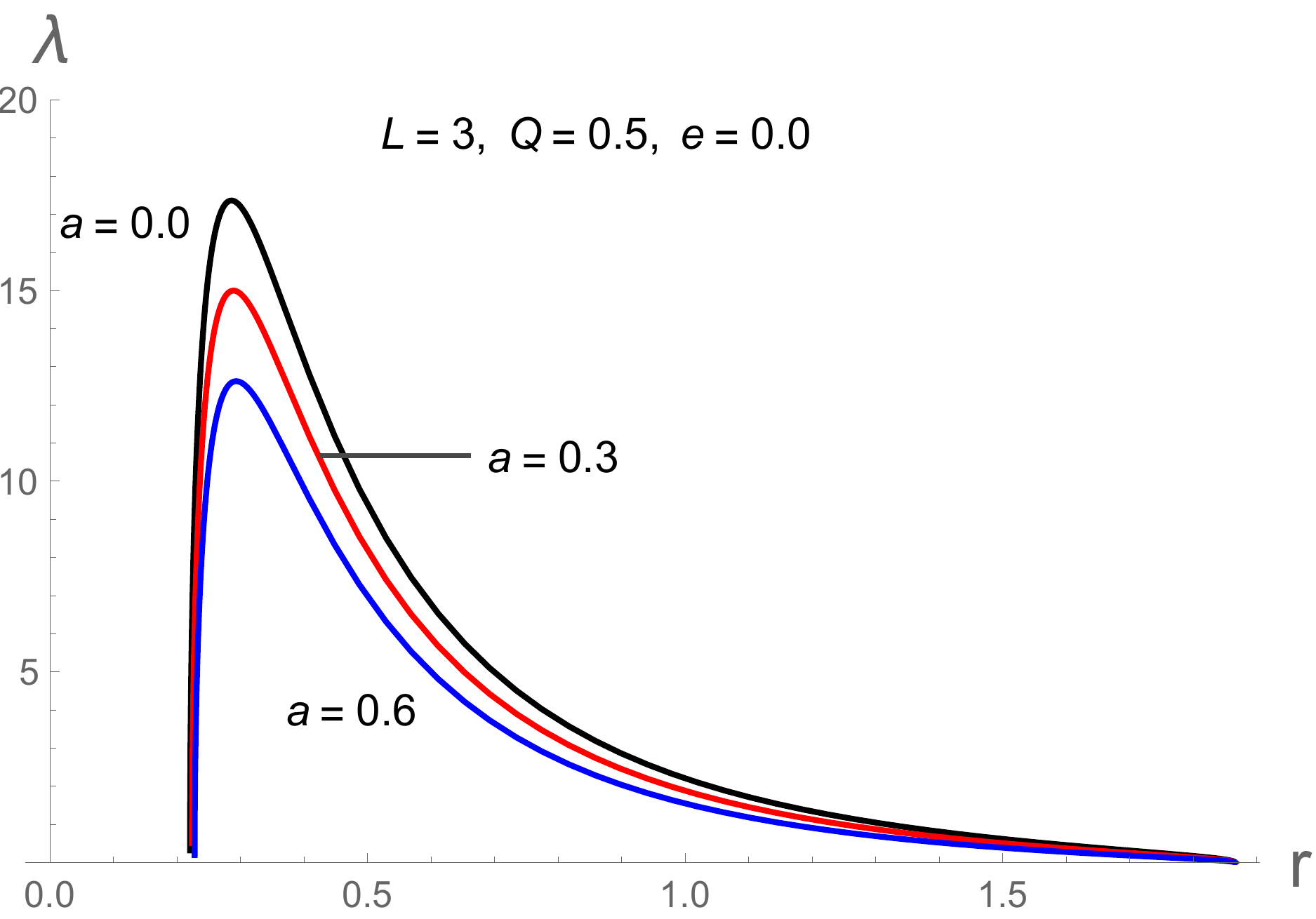}
    \end{minipage}
    \vspace{0.5cm}
        \begin{minipage}[b]{0.58\textwidth} \hspace{-1.1cm}
       \includegraphics[width=.7\textwidth]{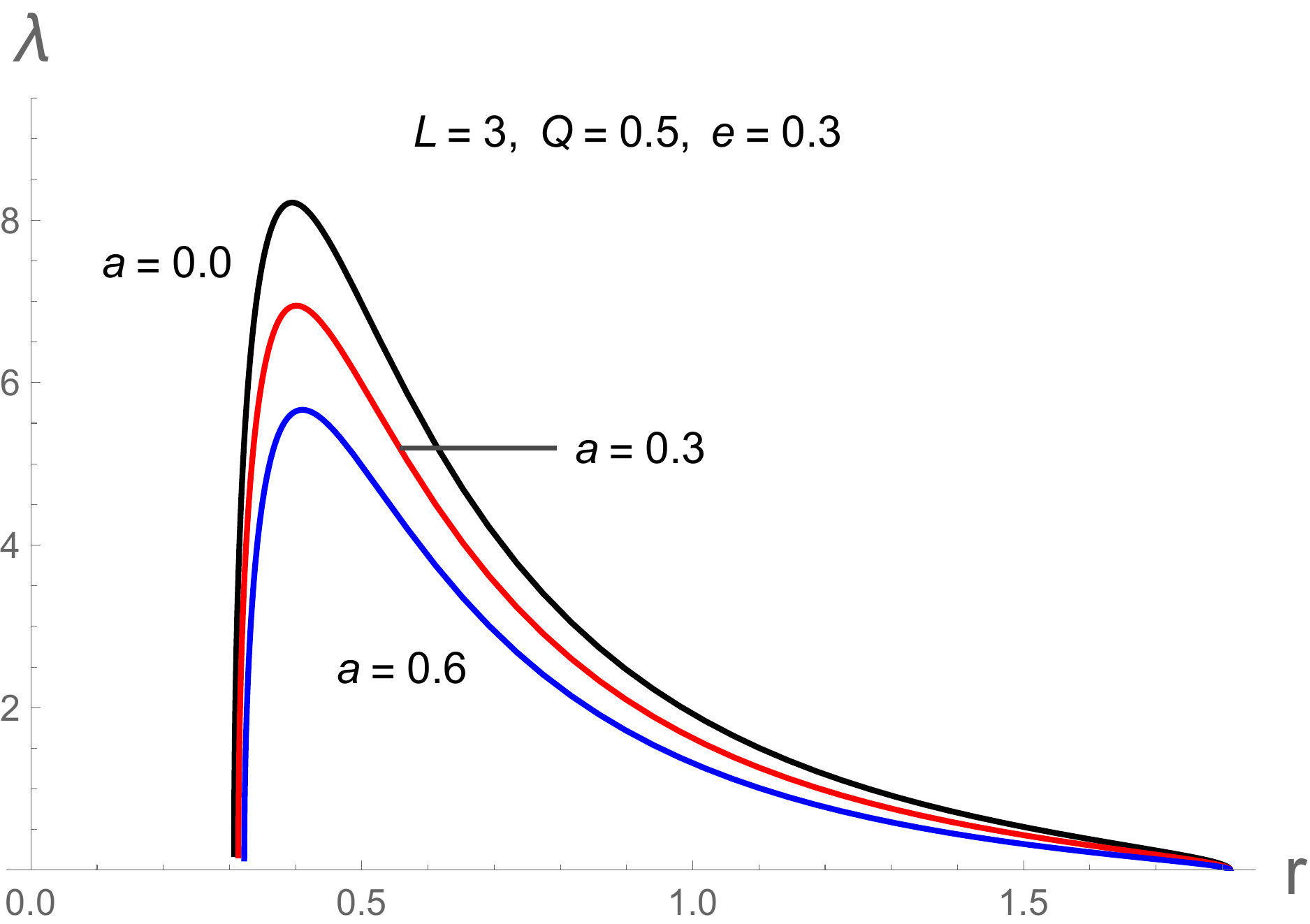}
    \end{minipage}
\begin{center}
\begin{minipage}[b]{0.58\textwidth} \hspace{1.9cm}
        \includegraphics[width=0.7\textwidth]{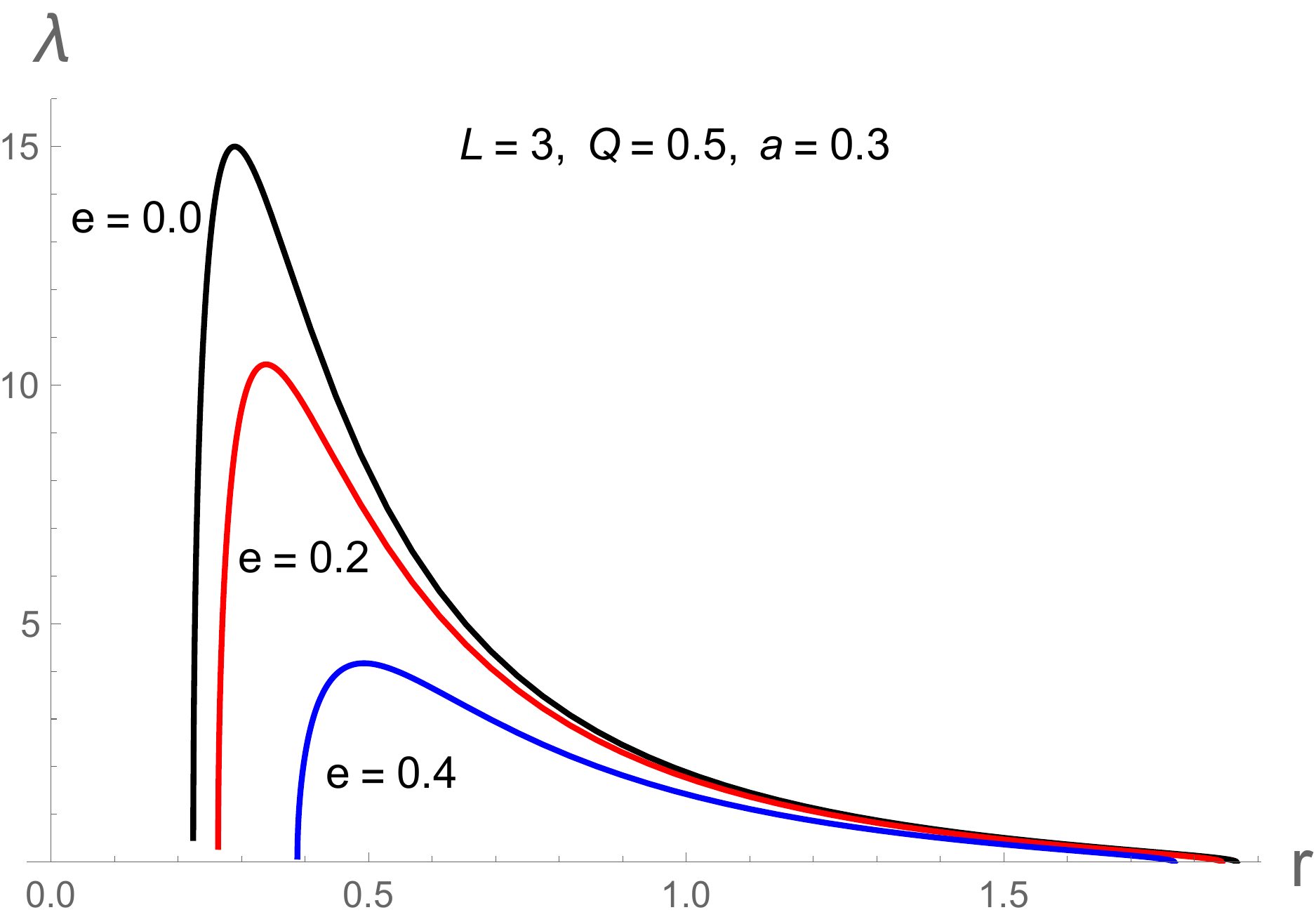}
    \end{minipage}
\end{center}
\caption{Graphical description of the Lyapunov exponent $\lambda$ along radial distance $r$, with the assumption of $\mathcal{E}=1$.}\label{Lexponent}
\end{figure*}
The radial profile of the Lyapunov exponent is described in Fig. \ref{Lexponent}. The upper row is plotted for KN and KNK BH, respectively in the left and right channels for the variations of $a$. It is noted that both BH rotation and charge decreases the instability of circular orbits, while the KNK BH have less unstable orbits, as compared to the KN BH. Moreover, in the lower plot, it is shown that BH electric charge $e$, also decrease the instability of circular orbits and the nearby orbits converge at a larger radial distance $r$.
\section{The Penrose mechanism}
\label{sec:4}
In this section, we will explore the typical Penrose mechanism for the efficiency of energy extraction within ergoregion of a KNK BH spacetime. It is among the engaging topic of general relativity and its principal goal is to extract energy from a rotating BH. The mechanism itself depends on the conservation of energy, as well as momentum. It is a more efficient and rigorous technique for extraction of energy, as compared to the techniques of nuclear reactions. The necessary and sufficient conditions for the energy extraction from a rotating BH, via Penrose mechanism is the absorption of angular momentum and negative energy. Our primary goal is to figure out the effects of rotation $a$ and the dyonic charge on the efficiency of energy extraction.
\subsection{Negative energy states}
Negative energy states within ergosphere of a BH has essential consequences in BH physics. They could be occurred due to the counter-rotating orbits (as in Kerr BH), as well as due to the electromagnetic interaction (as in RN BH) \cite{Shahzadi}. Finding the energy limits of a particle is of considerable interest, which they have at a particular position. On the equatorial plane, the radial equation \eqref{BS8}, can be rewritten as
\bea \label{negenergy}
\mathcal{E}^{2}\left[(r^{2}+a^{2})r^2 +a^2(2Mr-e{^2}-Q{^2})\right] - 2a\mathcal{E}L(2Mr-e{^2}-Q{^2})+L^{2}(2Mr-r^2-e{^2}-Q{^2})+r^2 \Delta{m_0^2}=0.
\eea
On solving the above equation for $\mathcal{E}$ and $L$, one can obtain
\beq\label{NE0}
\mathcal{E}=\frac{aL(2Mr-(e{^2}-Q{^2}) \pm \mathcal{Z}_{1}
\sqrt{\Delta}} {r^{4}+a^{2}(2Mr+r^2-e{^2}-Q{^2})},
\eeq
\beq\label{NE1}
L=\frac{-a\mathcal{E}(2Mr-(e{^2}-Q{^2})\pm \mathcal{Z}_{2}
\sqrt{\Delta}} {r^2-2Mr+e{^2}+Q{^2}},
\eeq
where
\beq\nonumber
\mathcal{Z}_{1}=\sqrt{L^{2}r^{4}-[r^{4}+a^{2}(2Mr+r^2-e{^2}-Q{^2})]r^2{m_0^2}}\,,
\eeq
\beq\nonumber
\mathcal{Z}_{2}=\sqrt{\mathcal{E}^{2}r^{4}+(r^2-2Mr+e{^2}+Q{^2})r^2{m_0^2}}\,.
\eeq
We have made use of the following identity to obtain the above results
\beq\label{NE2}
r^{4}\Delta - a^{2}(2Mr-e{^2}-Q{^2})^{2}=(r^{4}+a^{2} (r^{2}+2Mr-e{^2}-Q{^2}))(r^{2}-2Mr+e{^2}+Q{^2}).
\eeq
Equation \eqref{NE0} can be utilized to deduce the condition of negative energy states. Initially, we set an energy $\mathcal{E}=1$ and consider only $+$ sign of Eq. \eqref{NE0}. For $\mathcal{E}<0$, it also necessary that $L<0$, and
\bea\nonumber
\left[a^2 L^2(2Mr-e{^2}-Q{^2})\right]^{2}>\left[ L^{2}r^{2}-  \lbrace  r^{4}+a^{2}(r^{2}+2Mr-e{^2}-Q{^2}) \rbrace {m_0^2} \right]r^{2}\Delta.\label{NE3}
\eea
With the help of Eq. \eqref{NE2}, the above inequality \eqref{NE3}, can be rewritten as
\bea\nonumber
\left[(r^{2}-2Mr+e{^2}+Q{^2})L^{2}-{m_0^2} \Delta r^{2} \right] \left[a^{2}( r^{2}+2Mr-e{^2}-Q{^2})+r^{4}\right]<0.\label{NE4}
\eea
From the above inequality, it can be shown that $\mathcal{E}<0 $, if and only if $L<0$,
and
\begin{eqnarray}
\frac{r^2-2Mr+e{^2}+Q{^2}}{r^2}<\frac{\Delta}{L^{2}}{{m_0^2}}.
\end{eqnarray}
Thus we may conclude that only counter-rotating particles can have negative energy; and on the equatorial plane, it is further necessary that the particles must be inside the ergosphere. On substituting both $e=Q=0$, the above inequality reduces to the case of Kerr BH \cite{Chandrasekhar}.
\subsection{Wald inequality}
Investigating energy limits for the Penrose energy extraction process is of key interest. For this purpose, Wald \cite{Wald} introduced an inequality that can be used to find limits of the Penrose energy extraction process. To find out these limits on the KNK BH, we assume a particle of specific energy $\mathcal{E}$ and four velocity $U^{\sigma}$, splits into fragments. Let $\mathcal{\varepsilon}$ be the specific energy and $u^{\sigma}$ be the four-velocity of one of the fragments. For obtaining the limits  of $\mathcal{\varepsilon}$, we assume that $e_{l}^{\sigma}$ be an orthonormal tetrad frame in which $U^{\sigma}$ coincides with $e_{0}^{\sigma}$ and the remaining basis vectors are $e_{(k)}^{\sigma} (k=1,2,3)$. In this frame
\beq \label{WI01}
u^{\sigma} = \rho (U^{\sigma} + \upsilon^{(k)} e_{(k)}^{\sigma}),
\eeq
here $\upsilon^{(k)}$ represents the spatial components of the three velocity of the fragment
$\rho=1 / \sqrt{1 - |\upsilon |^{2}}$ with\\ $|\upsilon |^{2} = \upsilon^{(k)}\upsilon_{(k)}$. Since spacetime allows the timelike Killing vector $\zeta_{\sigma}=\partial / \partial x^{0}$, it can be represented in the tetrad frame as
\beq\label{WI02}
\zeta_{\sigma}=\zeta_{(0)}U_{\sigma} + \zeta_{(i)}e^{(k)}_{\sigma}.
\eeq
Henceforth, in terms of the Killing vector $\mathcal{E}$ can be expressed as
\beq\label{WI03}
\mathcal{E} = \zeta_{\sigma} U^{\sigma} = \zeta_{(0)} = \zeta^{\sigma}U_{\sigma} = \zeta^{(0)},
\eeq
and
\beq\label{WI04}
g_{00} = \zeta^{\sigma} \zeta_{\sigma} = -\zeta_{(0)}^{2} + \zeta_{(k)} \zeta^{(k)} = -\mathcal{E}^{2} + |\zeta|^{2}.
\eeq
Consequently, we get
\beq\label{WI05}
|\zeta|^{2} = \zeta_{(k)} \zeta^{(k)} = \mathcal{E}^2+g_{00}.
\eeq
From Eq. \eqref{WI01}, we have
\beq\label{WI06}
\mathcal{\varepsilon} = \zeta_{(\sigma)}u^{(\sigma)} = \rho(\zeta_{(0)} + \upsilon^{(k)} \zeta_{(k)}) = \rho(\mathcal{E} + |\upsilon| |\zeta| \cos\theta),
\eeq
in which $\theta$ is the angle between $\upsilon^{(k)}$ and $\zeta_{(k)}$. By making use of Eq. \eqref{WI05}, the Eq. \eqref{WI06} simplifies to
\beq\label{WI07}
\mathcal{\varepsilon} = \rho \mathcal{E} + \rho |\upsilon| \sqrt{\mathcal{E}^{2}+g_{00}}\cos\theta.
\eeq
Equation \eqref{WI07}, can be written as
\beq\label{WI08}
\rho \mathcal{E} - \rho |\upsilon| \sqrt{\mathcal{E}^{2}+g_{00}} \leq \mathcal{\varepsilon} \leq \rho \mathcal{E} + \rho |\upsilon| \sqrt{\mathcal{E}^{2}+g_{00}}.
\eeq
In the case of KNK BH, the Wald inequality takes the form
\beq\label{WI09}
\rho \mathcal{E} - \rho |\upsilon| \sqrt{\mathcal{E}^{2}+1-e^2-Q^2} \leq \mathcal{\varepsilon} \leq \rho \mathcal{E} + \rho |\upsilon| \sqrt{\mathcal{E}^{2}+1-e^2-Q^2}.
\eeq
By following \cite{Prad2}, the maximum energy of a particle revolving in a stable circular orbit can be found as
\beq\label{WI10}
\mathcal{E}_{0}=\frac{1}{\sqrt{3 - e^2-Q^2}}.
\eeq
In case of $\mathcal{\varepsilon}<0$, it is necessary that
\beq\label{WI11}
|\upsilon| > \frac{ \mathcal{E}}{\sqrt{\mathcal{E}^{2}+1-e^2-Q^2}}=\frac{1}{2-e^2-Q^2}.
\eeq
On the contrary, the fragments could have relativistic energies which may be possible before the energy extraction. On substituting $e=Q=0$, Eq. \eqref{WI11} will reduce to the case of Kerr BH \cite{Chandrasekhar}.
\subsection{The efficiency of Penrose mechanism}
The problem of energy extraction and its efficiency via the Penrose mechanism from a revolving BH is among the topics of plentiful interest in BH energetics. Initially, we assume a particle consisting energy $E_{(0)}$, split into two sub-particles specifically 1 and 2, (respectively posses energy $E_{(1)}$ and $E_{(2)}$) after entering into the ergoregion of a BH. In comparison with the incident particle, the sub-particle 1 has more energy and thus leave the ergoregion whereas, the sub-particle 2 possess negative energy falls into the BH. By making use of the law of energy conservation
\beq \nonumber
\mathcal{E}_{(0)}=\mathcal{E}_{(1)}+\mathcal{E}_{(2)}.
\eeq
In the above equation $\mathcal{E}_{(2)}<0$, provides that $\mathcal{E}_{(1)} > \mathcal{E}_{(0)}$. Let us assume that with respect to an observer at infinity particles radial velocity  be $\vartheta=dr/dt$. 
Henceforth, utilizing the laws of conservation of energy and angular momentum
\beq\label{Eff1}
\mathcal{E}=-p^{t}\mathcal{X}, \quad L=p^{t}\Omega,
\eeq
with
\beq
\mathcal{X} \equiv g_{tt} + g_{t\phi}\Omega.
\eeq
Using $p^{\nu} p_{\nu}=-m^{2}$, we have
\beq\label{Eff2}
g_{tt}\dot{t}^{2}+2g_{t\phi}\dot{t}\dot{\phi}+g_{rr}\dot{r}^{2}+g_{\phi\phi}\dot{\phi}^{2}=-m^{2}.
\eeq
Dividing the above equation by $\dot{t}^{2}$, result in
\beq\label{Eff3}
g_{tt}+2 \Omega\ g_{t\phi}+\Omega^{2}g_{\phi\phi} +\frac
{\vartheta^{2}}{\Delta}r^{2}=-\left(\frac{m
\mathcal{X}}{\mathcal{E}}\right)^{2}.
\eeq
In the above equation, the last term of the left-hand side is always positive, while the right-hand side is negative or maybe zero. Consequently, the above equation could be expressed by
\beq\label{Eff4}
g_{tt}+2 \Omega\ g_{t\phi}+\Omega^{2}g_{\phi\phi}=-\left(\frac{m
\mathcal{X}}{\mathcal{E}}\right)^{2}-\frac {\vartheta^{2}}{\Delta}r^{2}\leq 0.
\eeq
By making use of Eq. \eqref{Eff1}, the equations of conservation of energy and angular momentum takes the form
\bea\label{Eff5}
p^{t}_{(0)}\Omega_{(0)}=p^{t}_{(1)}\Omega_{(1)}+p^{t}_{(2)}\Omega_{(2)},
\eea
\beq\label{Eff6}
p^{t}_{(0)}\mathcal{X}_{(0)}=p^{t}_{(1)}\mathcal{X}_{(1)}+p^{t}_{(2)}\mathcal{X}_{(2)}.
\eeq
As a result, the efficiency of Penrose mechanism could be expressed by
\beq
\eta=\frac{\mathcal{E}_{(1)}-\mathcal{E}_{(0)}}{\mathcal{E}_{(0)}}=\chi-1.
\eeq
In the above equation $\chi=\mathcal{E}_{(1)}/\mathcal{E}_{(0)}>1$, while using Eqs. \eqref{Eff1},
\eqref{Eff5} and \eqref{Eff6}, we obtained
\beq\label{Eff7}
\chi=\frac{\mathcal{E}_{(1)}}{\mathcal{E}_{(0)}}=\frac{(\Omega_{(0)}-\Omega_{(2)})\mathcal{X}_{(1)}}
{(\Omega_{(1)}-\Omega_{(2)})\mathcal{X}_{(0)}}.
\eeq
Next, we consider an incident particle possess energy $\mathcal{E}_{(0)}=1$ goes into the ergoregion and split into two photons of momenta $p^{(1)}=p^{(2)}=0$. Moreover, from the above equation, it could be noted that the efficiency can be maximized by choosing the smallest value of $\Omega_{(1)}$, while the largest value of $\Omega_{(2)}$ at the same time, it requires $\vartheta_{(1)}=\vartheta_{(2)}=0$. In this case
\beq\label{Eff8}
\Omega_{(1)}=\Omega_{+}, \quad \Omega_{(2)}=\Omega_{-}.
\eeq
The corresponding values of parameter $\mathcal{X}$ are
\beq\label{Eff9}
\mathcal{X}_{(0)}=g_{tt}+\Omega_{(0)} \ g_{t\phi}, \quad
\mathcal{X}_{(2)}=g_{tt}+\Omega_{-} \ g_{t\phi}.
\eeq
The four-momenta of pieces are
\beq\nonumber
p_{\nu}=p^{t}(1,0,0,\Omega_{\nu}),\quad \nu=1,2.
\eeq
Henceforth, Eq. \eqref{Eff2} simplifies to
\bea
(g_{t\phi}^{2}+g_{\phi\phi})\Omega^{2} + 2\Omega(1+g_{tt})g_{t\phi}+(1+g_{tt})g_{tt}=0.
\eea
Utilizing the above equation, angular velocity of the original particle will modify to
\beq
\Omega_{(0)}=\frac{-(1+g_{tt})g_{t\phi}+\sqrt{(1+g_{tt})(g^{2}_{t\phi}-g_{\phi\phi}g_{tt})}}{g^{2}_{t\phi}+g_{\phi\phi}}.
\eeq
By inserting the values of Eqs. \eqref{Eff8} and \eqref{Eff9} into \eqref{Eff7}, efficiency of the Penrose mechanism can be acquired as
\beq\label{eta}
\eta=\frac{(g_{tt}+g_{t\phi} \Omega_{+})(\Omega_{(0)}-\Omega_{-})}
{(g_{tt}+g_{t\phi}\Omega_{0})(\Omega_{(+)}-\Omega_{-})}-1.
\eeq
In lights of the Penrose mechanism, the maximum efficiency ($\eta_{max}$) could only be acquired, if the incident particle split at the event horizon of a BH. Consequently, Eq. \eqref{eta} modifies to
\beq
\eta_{max}=\frac{1}{2}\left(\sqrt{\frac{a^2+r^2-\Delta}{r^{2}}}-1\right)\vert_{r=r_{+}}.
\eeq
\begin{table*}\caption{Table of $\eta_{max} (\%)$ from a KNK BH spacetime, via the Penrose mechanism.}
\begin{tabular}{cc c c c c c c c c c }
\hline \hline \noalign{\smallskip\smallskip}
& & {a=0.3}  &  {a=0.5}  &  {a=0.7}  &   {a=0.8}  &  {a=0.9}  &  {a=0.99}  &  {a=1.0}& \\ \noalign{\smallskip}
\hline \noalign{\smallskip\smallskip}
\multicolumn{1}{  c  }{\multirow{6}{*}{$Q=0$} } &
\multicolumn{1}{  c  }{ $e=0.0$ } & 0.5859 & 1.7638 & 4.00842 & 5.9017 & 9.00984 & 16.1956 & 20.7107 &    \\ 
\multicolumn{1}{  c  }{}                        &
\multicolumn{1}{  c  }{ $e=0.05$ } & 0.5867 & 1.7665 & 4.0163 & 5.9163 & 9.0433 & 16.4262 &  & \\
\multicolumn{1}{  c  }{}                        &
\multicolumn{1}{  c  }{ $e=0.1$ } & 0.5891 & 1.7746 & 4.0403 & 5.9608 & 9.1461 & 17.2818&    \\
\multicolumn{1}{  c  }{}                        &
\multicolumn{1}{  c  }{ $e=0.2$ }& 0.5988  & 1.8081 & 4.1403 & 6.1489 & 9.6001 & &\\
\multicolumn{1}{  c  }{}                        &
\multicolumn{1}{  c  }{ $e=0.3$ } & 0.6159 & 1.8678 & 4.3232 & 6.5055 & 10.5711 &  &  & \\
\multicolumn{1}{  c  }{}                        &
\multicolumn{1}{  c  }{ $e=0.5$ } & 0.6803  & 2.1005 & 5.1120 & 8.3289 &  &  &  &    \\ 
\hline \noalign{\smallskip\smallskip}

\multicolumn{1}{  c  }{\multirow{6}{*}{$Q=0.1$} } &
\multicolumn{1}{  c  }{ $e=0.0$ } & 0.5891  & 1.7746 & 4.0403 & 5.9608 & 9.1461 & 17.2818 &  &    \\ 
\multicolumn{1}{  c  }{}                        &
\multicolumn{1}{  c  }{ $e=0.05$ } & 0.5898  & 1.7774 & 4.0484 & 5.9758 & 9.1812 & 17.6569 &  &    \\
\multicolumn{1}{  c  }{}                        &
\multicolumn{1}{  c  }{ $e=0.1$ }& 0.5922 & 1.7856 & 4.0729 & 6.0216 & 9.2894 &  &  & \\
\multicolumn{1}{  c  }{}                        &
\multicolumn{1}{  c  }{ $e=0.2$ }& 0.6021 & 1.8197 & 4.1751 & 6.2156 & 9.7694 &  &  & \\
\multicolumn{1}{  c  }{}                        &
\multicolumn{1}{  c  }{ $e=0.3$ }& 0.6194  & 1.8804 & 4.3625 & 6.5846 & 10.813 &  &  & \\
\multicolumn{1}{  c  }{}                        &
\multicolumn{1}{  c  }{ $e=0.5$ }& 0.6850 & 2.1178 & 5.1765 & 8.5111 &  &  &  & \\
\hline \noalign{\smallskip\smallskip}

\multicolumn{1}{  c  }{\multirow{6}{*}{$Q=0.3$} } &
\multicolumn{1}{  c  }{ $e=0.0$ } & 0.6159 & 1.8678 & 4.3232 & 6.5055 & 10.5711 &  &  &    \\ 
\multicolumn{1}{  c  }{}                        &
\multicolumn{1}{  c  }{ $e=0.05$ } & 0.6168 & 1.8709 & 4.3329 & 6.5250 & 10.6297 &  &  &    \\
\multicolumn{1}{  c  }{}                        &
\multicolumn{1}{  c  }{ $e=0.1$ }& 0.6194 & 1.8804 & 4.3625 & 6.5846 & 10.8130 &  &  & \\
\multicolumn{1}{  c  }{}                        &
\multicolumn{1}{  c  }{ $e=0.2$ }& 0.63048 & 1.9195 & 4.4870 & 6.8408 & 11.6971 &  &  & \\
\multicolumn{1}{  c  }{}                        &
\multicolumn{1}{  c  }{ $e=0.3$ }& 0.6501 & 1.9897 & 4.7190  & 7.3476 & 14.6030 &  &  & \\
\multicolumn{1}{  c  }{}                        &
\multicolumn{1}{  c  }{ $e=0.5$ }& 0.7253 & 2.2713 & 5.8046 & 11.0580 &  &  &  & \\
\hline \hline \noalign{\smallskip}
\end{tabular}\label{Tab3}
\end{table*}
\begin{figure*}
 \begin{minipage}[b]{0.58\textwidth} \hspace{0.3cm}
        \includegraphics[width=0.75\textwidth]{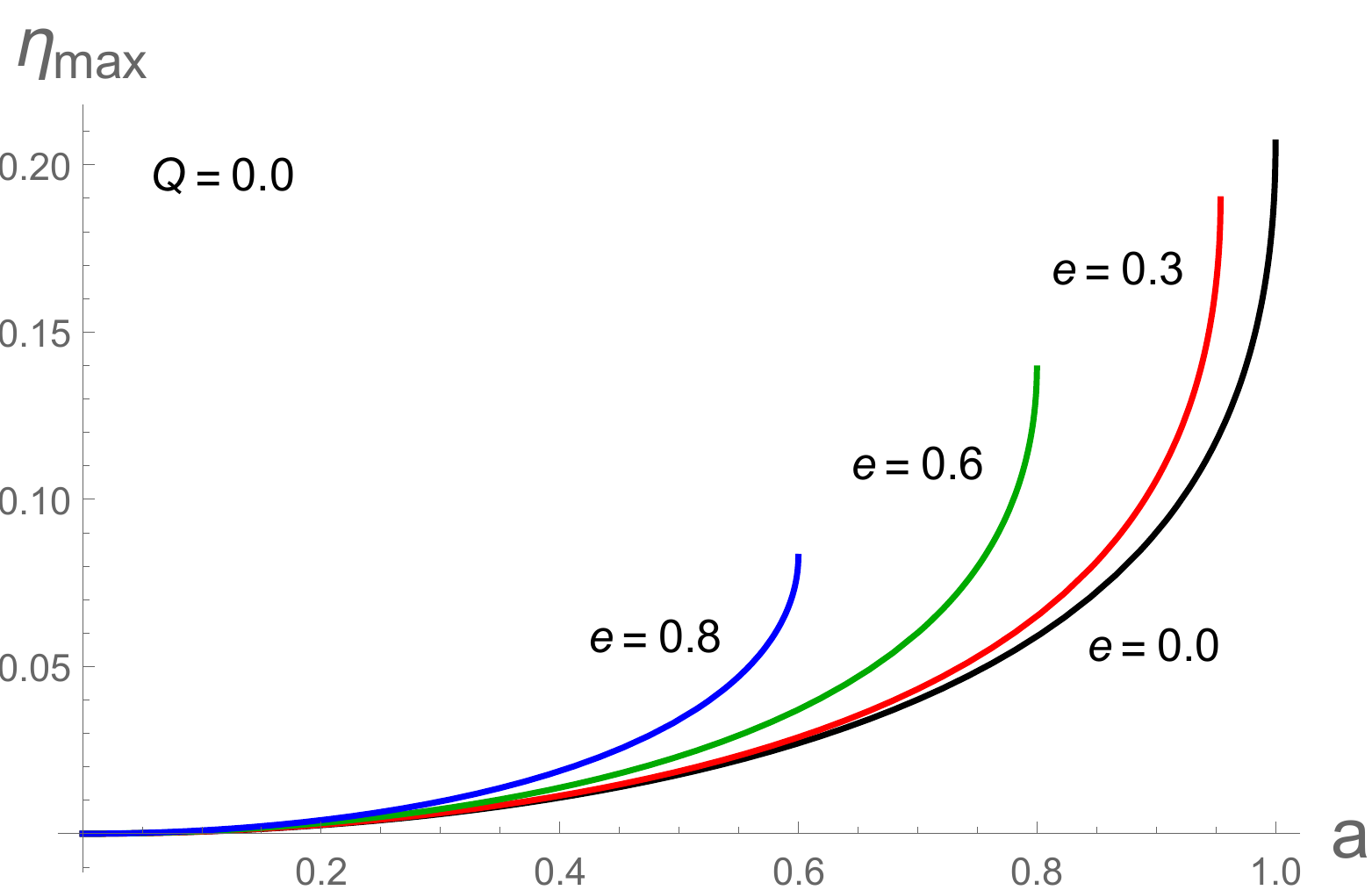}
    \end{minipage}
    \vspace{0.3cm}
        \begin{minipage}[b]{0.58\textwidth} \hspace{-1.1cm}
       \includegraphics[width=.75\textwidth]{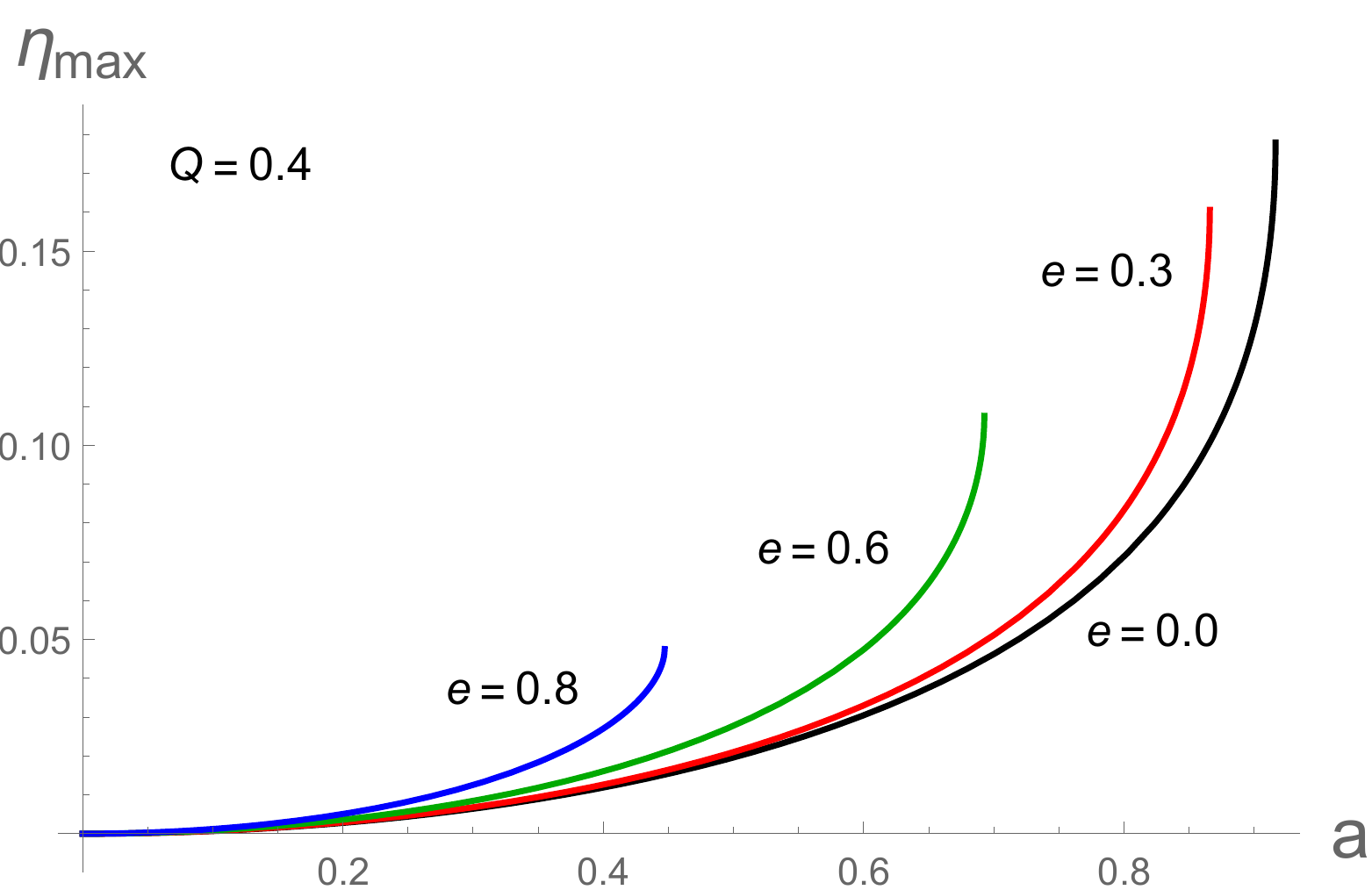}
    \end{minipage}
\caption{Graphical description of $\eta_{max}$, along with spin $a$ of the BH.}\label{EE1}
\end{figure*}
Table \ref{Tab3}, showing the numerical calculation of $\eta_{max}$, at various discrete values of spin and dyonic charge parameters of the KNK BH. The obtained results reveal that both BH rotation and the dyonic charge, contribute to the maximum efficiency of energy extraction. Moreover, it can also be observed from Table \ref{Tab3}, as well as Fig. \ref{EE1}, that energy extraction from the KNK BH, is higher in comparison with the KN BH in case of slow rotation. But the maximum value of $\eta_{max}$ can only be acquired at $a = 1$, which is possible only at $e = Q = 0$ (extreme Kerr BH).
While it can also be observed that in the case of fast-rotating BH, energy extraction increases along with BH spin $a$.
\section{Conclusion}
\label{sec:5}
In current work, we have studied particle dynamics of massless particles together with the Penrose mechanism around a KNK BH spacetime. The event and Cauchy horizons, as well as both of the inner and outer ergosurfaces, are investigated at various discrete points of spin and dyonic charge of the BH. Moreover, it is found that BH rotation and the dyonic charge contributes to both the Cauchy horizon and inner ergosurface, whereas diminishing both of the event horizon and static limit. Geodesics are essential to investigate particle dynamics around a BH, but among various kinds of geodesics, the circular one is much engaging. Since, the circular geodesics can be used to investigate the motion of galaxies, as well as the theory of accretion and planetary dynamics.

Using the equations of motion, we have obtained expressions for the radii of photon orbits and the MSCOs. It is observed that BH charge decreases the radii of photon orbits and the MSCOs for both prograde and retrograde particle motions. On the other hand, its spin diminishing the radii of both photon orbits, as well as the MSCOs of prograde particle motions, while contributes to the radii of retrograde particle motions. To analyse the stability of circular orbits, we have studied the effective potential and Lyapunov exponent. The existence of stable and unstable circular orbits are, respectively analyzed with the help of relative minima and maxima of the effective potential. We have examined that both spin and dyonic charge, decreases the stability of circular orbits. Furthermore, BH spin and dyonic charge, resulting in decreasing the instability of the Lyapunov exponent and its instability also decrease with the increase of radial distance $r$.

The process of energy extraction around KNK spacetime has been examined via Penrose mechanism. We have discussed the negative energy states, Wald inequality, as well as the efficiency of energy gain during the Penrose process and shown that negative energy can occur, if and only if the angular momentum is negative. It is concluded that both BH rotation and dyonic charge contributes to the process of energy extraction. In the presence of charge and BH rotation, particles get more energy which helps to extract more energy from the BH. It is shown that along with the radial distance $r$, the efficiency of energy gain  $\eta$ increase with the increase of BH rotation. For the extreme Kerr BH ($a=1, e=Q=0$), $\eta_{max}=20.7\%$ \cite{Chandrasekhar}. Besides, our acquired results are considerably identical to that of Kerr BH surrounded by a magnetic field, as in both cases energy gain is remarkably affected by the BH rotation \cite{Dadhich}. It is concluded that one can obtain more energy via the Penrose mechanism in case of KNK BH, as compared to the KN and Kerr BHs.
\subsubsection*{Acknowledgment}
This work is supported by the NSFC Project (11771407) and the MOST Innovation Method Project (2019IM050400).
\bibliographystyle{unsrt}

\end{document}